\documentclass[12pt]{iopart}

\usepackage{iopams,setstack,graphicx,here}
\usepackage{cite}
\eqnobysec

\usepackage{color}

\newcommand{\mean}[1]{\left < #1 \right >}

\newcommand{\abs}[1]{\left | #1 \right |}

\newcommand{\Tp}[0]{T_{\mbox{\tiny{$\parallel$}}}}
\newcommand{\Ts}[0]{T_{\perp}}

\newcommand{\eqnlabel}[1]{\refstepcounter{equation}\label{#1}\addtocounter{equation}{-1}}
\newcommand{\eqnref}[1]{(\ref{#1})}

\newcounter{Aeqnval}
\def\Anumparts{\addtocounter{equation}{1}%
     \setcounter{Aeqnval}{\value{equation}}%
     \setcounter{equation}{0}%
     \def\theequation{\ifnumbysec
     \Alph{section}.\arabic{Aeqnval}{\it\alph{equation}}%
     \else\arabic{Aeqnval}{\it\alph{equation}}\fi}}

\def\endAnumparts{\def\theequation{\ifnumbysec
     \Alph{section}.\arabic{equation}\else
     \arabic{equation}\fi}%
     \setcounter{equation}{\value{Aeqnval}}}
\begin{document}

	%%%%%%%%%%%%%%%%%%%%%%%%%%%%%%%%%%%%%%%%%%%%%%%%%%%%%%%%%%%%%%%%%%%%%%%%%%%%%%
	%%								TITLEPAGE

	%title, auther, adress, mail
	\title[Active Brownian particles with velocity-alignment and active fluctuations]{Active Brownian particles with velocity-alignment and active fluctuations}
	\author{R Gro{\ss}mann$^1$, L Schimansky-Geier$^1$ and P Romanczuk$^{2}$}

	\address{$^1$ Department of Physics, Humboldt Universit\"at zu Berlin, Newtonstr. 15, 12489 Berlin, Germany}
	\address{$^2$ Max Planck Institute for the Physics of Complex Systems, N\"othnitzerstr. 38, 01187 Dresden, Germany}

	\ead{prom@pks.mpg.de}

	%abstract
	\begin{abstract}	
We consider a model of active Brownian particles with velocity-alignment in two spatial dimensions with passive and active fluctuations. Hereby, active fluctuations refers to purely non-equilibrium stochastic forces correlated with the heading of an individual active particle. In the simplest case studied here, they are assumed as independent stochastic forces parallel (speed noise) and perpendicular (angular noise) to the velocity of the particle. 
On the other hand, passive fluctuations are defined by a noise vector independent of the direction of motion of a particle, and may account for example for thermal fluctuations.

We derive a macroscopic description of the active Brownian particle gas with velocity-alignment interaction. Hereby, we start from the individual based description in terms of stochastic differential equations (Langevin equations) and derive equations of motion for the coarse grained kinetic variables (density, velocity and temperature) via a moment expansion of the corresponding probability density function. 

We focus here in particular on the different impact of active and passive fluctuations on the onset of collective motion and show how active fluctuations in the active Brownian dynamics can change the phase-transition behaviour of the system. In particular, we show that active angular fluctuation lead to an earlier breakdown of collective motion and to emergence of a new bistable regime in the mean-field case.
	\end{abstract}	
		
	
	%\submitto{\NJP}

	\maketitle
	%%%%%%%%%%%%%%%%%%%%%%%%%%%%%%%%%%%%%%%%%%%%%%%%%%%%%%%%%%%%%%%%%%%%%%%%%%%%%%

	%%%%%%%%%%%%%%%%%%%%%%%%%%%%%%%%%%%%%%%%%%%%%%%%%%%%%%%%%%%%%%%%%%%%%%%%%%%%%%
	%%		
	%									MAIN

	\section{Introduction}
	The fascinating self-organisation phenomenon of collective motion in biology, such as exhibited by flocks of birds or schools of fish, has attracted scientists from very different backgrounds over the past decades. Whereas biologists focus on the evolutionary advantages and the biological function of moving in groups \cite{couzin_collective_2002,krause_living_2002,parrish_complexity_1999}, physicist address rather the question about universal laws, governing the collective dynamics either from the theoretical point of view \cite{vicsek_novel_1995,toner_long-range_1995,toner_flocks_1998,gregoire_onset_2004,aldana_phase_2007,chate_collective_2008,romanczuk_collective_2009} or, as done recently, by analysing certain model systems experimentally \cite{kudrolli_swarming_2008,deseigne_collective_2010,schaller_polar_2010,peruani_collective_2012}. In general, physicist are inclined towards the analysis of simple, but generic, models of collective motion such as the minimal model proposed by Vicsek \etal \cite{vicsek_novel_1995}. The Vicsek model can be considered as a non-equilibrium extension of the XY-model, where self-propelled particles move with a finite speed in continuous space and at discrete times align their velocities with the average velocity of other particles sensed in their local neighbourhood in the presence of noise. The Vicsek model exhibits a non-equilibrium phase transition leading to the onset of long range order at sufficiently low noise and high particle density.  Detailed studies of the Vicsek model have led to an intense debate on the nature of the phase transition (continuous vs. discontinuous) \cite{gregoire_onset_2004,nagy_new_2007,aldana_phase_2007,chate_collective_2008,baglietto_nature_2009}. 

One of the main characteristics of the Vicsek model, as well as of many other individual-based models of collective motion, is the assumption of a constant speed of individuals. In general, this does not hold for the motion of real biological agents \cite{bazazi_nutritional_2011}. Furthermore, recent experimental realizations of swarming in non-living systems are based on rectification of fluctuation due to asymmetric friction and therefore imply a non-negligible role of speed fluctuations \cite{kudrolli_swarming_2008,deseigne_collective_2010}.  

A possibility to introduce speed fluctuations into the dynamics of active particles is the addition of random increments from a well-defined probability distribution with zero mean to the speed of an individual in numerical simulations \cite{peruani_self-propelled_2007,stroembom_collective_2011}. This approach is useful to analyze the robustness of the collective dynamics with respect to noise. However, in this kind of models the speed dynamics is always independent of the interactions between individuals. 

A different approach is based on the formulation of stochastic equations of motion for the positions ${\bi r}_i$ and velocities ${\bi v}_i$ of $N$ interacting individuals ($i=1,\dots,N$):
	\numparts
	\begin{eqnarray}
		&\frac{\rmd \bi{r}_i}{\rmd t}   =\bi{v}_i \, \mbox{, } \\
		m_i \, &\frac{\rmd \bi{v}_i}{\rmd t}=\bi{F}_i + \boldsymbol{\eta}_i.	\label{eq:eom_gen}
	\end{eqnarray} 
	\endnumparts
Here, $m_i$ is the mass of the focal individual, 
$\bi{F}_i$ is an effective force vector governing the deterministic dynamics, whereas $\boldsymbol{\eta}_i$ models an effective stochastic force, which account for the randomness in the motion of the individual.
In general, both terms can depend on the position and velocities of all other individuals. In addition, the stochastic force $\boldsymbol{\eta}_i$, must, by definition, explicitly depend on time, which is not necessarily the case for the deterministic force $\bi{F}_i$.

From a physical point of view, the velocity equation (\ref{eq:eom_gen}) can be considered as a Langevin equation with actual physical forces on the right hand side. However, the full description of all physical forces determining the motion of an individual will be extremely complex and not very practical in most cases. Thus from a more general point of view, we can consider $\bi{F}_i$ and $\boldsymbol{\eta}_i$ as effective forces responsible for accelerations of the focal individual in response to external stimuli and/or due to internal decision processes. The functional form of $\bi{F}_i$ and $\boldsymbol{\eta}_i$ can be set using reasonable modeling assumptions based on qualitative empirical observations \cite{couzin_collective_2002,romanczuk_collective_2009}, or, if available, directly on experimental measurements of interactions between individuals \cite{lukeman_inferring_2010,katz_inferring_2011,herbert-read_inferring_2011}.

This class of variable speed models not only overcomes the problems mentioned above, but also allows a straight forward mathematical analysis of the dynamical behaviour due to the usage of stochastic differential equations. We refer to active particles with self-propulsion, where both, the magnitude and the direction of the velocity are (stochastic) degrees of freedom, as active Brownian particles (ABP). This allows a clear distinction from so-called self-propelled particles (SPP), which usually refers to active particles moving with constant speed \cite{vicsek_novel_1995,simha_hydrodynamic_2002}.
Originally, ABP referred to Brownian agents at far-from-equilibrium conditions with an internal energy depot, which can take-up energy from the environment and convert it into kinetic energy of motion \cite{schweitzer_complex_1998,erdmann_brownian_2000,schweitzer_brownian_2003,garcia_testing_2011}. However, as non-equilibrium conditions and local energy conversion are characteristic features for all active particle systems, we suggest the above general definition of ABP.

Recently, two of the authors considered non-interacting ABP with so-called passive and active fluctuations \cite{romanczuk_brownian_2011}. Passive fluctuations are assumed to be independent of the direction of motion, and may be motivated for example by a fluctuating environment. This could be e.g. thermal fluctuations, but in general also other non-equilibrium processes. Active fluctuations, on the other hand, are assumed to be correlated with the current direction of motion of the active particle. In the simplest case, they consist of independent stochastic processes parallel and perpendicular to the instantaneous direction of motion. Such fluctuations may be associated with fluctuations of the actual propulsion mechanism and are, per definition, a pure far-from-equilibrium phenomenon. 

The main focus of this work is the analysis of the impact of different fluctuation types on collective dynamics of interacting ABP. Here, we consider a generic, effective alignment interaction between individual particles, which represents a simple extension of the Vicsek model to the variable speed case. 
In particular, we will derive and analyze kinetic equations of ABP with velocity alignment and both passive and active fluctuations. Please note that already in the context of the Vicsek model, it was argued that different noise implementations may change the nature of the phase transition \cite{aldana_phase_2007,chate_comment_2007,baglietto_nature_2009}.

\section{Active Brownian particles with velocity-alignment}

\subsection{Microscopic description}

We describe the active Brownian particle gas by a system of coupled stochastic differential equations. The dynamics of the $i^{\mbox{\footnotesize{th}}}$ point-like particle ($i=1,...,N$) with mass $m_i$ is determined by the following set of Langevin equations interpreted in the sense of Stratonovich:
	\eqnlabel{eqn:ABM:active:particle:eqn}
	\numparts
		\begin{eqnarray}
			 & \frac{\rmd \bi{r}_i}{\rmd t}  &=  \bi{v}_i	\, \mbox{, }		\label{eqn:ABM:active:particle:eqna} \\
		     m_i \, & \frac{\rmd \bi{v}_i}{\rmd t}  &=  - \gamma(\bi{v}_i)\bi{v}_i + \tilde{\mu} \left (\bi{u}_{A_i} - \bi{v}_i  \right ) + \boldsymbol{\eta}_i(t) \, .												\label{eqn:ABM:active:particle:eqnb}
		\end{eqnarray}
	\endnumparts
%The position and velocity of the particle in two spatial dimensions is indicated by $\bi{r}_i(t)$ and $\bi{v}_i(t)$, respectively. For simplicity we assume that all particles have the same properties, especially the same mass: $m_i = m \; \forall \, i$. 

The self-propulsion of active particles is described by a velocity dependent friction coefficient $\gamma(\bi{v}_i)$, which can take both negative and positive values. Here, we use a piecewise linear function, the so-called Schienbein-Gruler friction \cite{schienbein_langevin_1993,erdmann_brownian_2000}: 
	\begin{eqnarray}
		\label{eqn:ABM:friction:function}
		-\gamma(\bi{v}_i)\bi{v}_i = -\alpha \left (1 - \frac{v_0}{\abs{\bi{v}_i}} \right )\bi{v}_i = \alpha (v_0 - \abs{\bi{v}_i})\bi{e}_{i,v}.
	\end{eqnarray}
The direction of motion of the $i^{\mbox{\footnotesize{th}}}$ particle is indicated by a unit vector $\bi{e}_{i,v}(t)=\bi{v}_i/ \!\abs{\bi{v}_i}$. Particles moving with small speeds $\abs{\bi{v}_i}<v_0$ accelerate into their direction of motion whereas their velocity is damped for high speeds $\abs{\bi{v}_i}>v_0$ leading to an average velocity unequal to zero. The relaxation time of the velocity dynamics is given by $\tau_{\alpha} = m \alpha^{-1}$ in the deterministic case. 
	
The above friction function is apolar in the sense that the particles do not distinguish between their front and back. A corresponding polar friction function with linear velocity relaxation was discussed by Romanczuk and Schimansky-Geier \cite{romanczuk_brownian_2011} and was used previously in various models of active particles (see e.g. \cite{condat_randomly_2005}). In the majority of cases where backward motion is negligible, e.g. in the case of moderate noise intensities, both models are equivalent. 

Due to the linear relaxation to the fixed point $v_0$, friction function \eref{eqn:ABM:friction:function} can be thought of as Taylor series of an arbitrary friction function around its stationary speed where $\gamma(v_0)=0$. 

The interaction of particles is considered to be a linear velocity-alignment mechanism with characteristic time $\tau_{\tilde{\mu}} = m \tilde{\mu}^{-1}$. In \eref{eqn:ABM:active:particle:eqnb}, the local mean velocity is denoted by
	\begin{eqnarray}
		\label{eqn:ABM:local:mean:velo}
		\bi{u}_{A_i} = \frac{1}{N_{A_i}} \, \sum_{j=1}^{N_{A_i}} \, \bi{v}_j \, \mbox{. }
	\end{eqnarray}
The focal particle $i$ aligns its velocity with the average velocity of all particles $N_{A_i}$ within a metric interaction region $A_{i}$ around its current position. In general, $A_{i}$ can have different geometries, however, a common choice in two spatial dimensions is a disc-like interaction zone with radius $\epsilon$.
The velocity-alignment strength is denoted by $\tilde{\mu}$. The summation in \eref{eqn:ABM:local:mean:velo} includes the $i^{\mbox{\footnotesize{th}}}$ particle itself so that in case of no neighbours ($N_{A_i}=1$) the alignment force vanishes, consistently. If the local mean velocity \eref{eqn:ABM:local:mean:velo} equals zero, the alignment interaction reduces to Stokes friction $\dot{\bi{v}}_i \propto - \tilde{\mu} \bi{v}_i$ decelerating the particles (``social friction''). 

In the limiting case of constant speeds $\abs{\bi{v}_i} \approx v_0$, the alignment interaction is related to the Kuramoto model \cite{kuramoto_self-entrainment_1975} describing synchronization phenomena of coupled oscillators. In this case, the polar angle defining the direction of motion corresponds to the phase of the oscillator \cite{paley_oscillator_2007,birnir_ode_2007}. Peruani \etal derived a mean-field theory for self propelled particles with constants speed and Kuramoto interaction \cite{peruani_mean-field_2008}. 

The above-mentioned velocity-alignment interaction has already been used to described collective dynamics, e.g. by Czir\'{o}k \etal \cite{czirok_formation_1996} in order to describe the formation of complex bacterial colonies as well as by Niwa \cite{niwa_self-organizing_1994} to characterize the dynamics of fish schools based on Langevin equations. 

In \eref{eqn:ABM:active:particle:eqnb}, $\boldsymbol{\eta}_i(t)$ denotes a stochastic force, introduced in \cite{romanczuk_brownian_2011}, contributing three terms to the velocity dynamics. The vector $\bi{e}_{i,\varphi}$ points perpendicular to the direction of motion and has unit length one. 
	\begin{eqnarray}
		\label{eqn:ABM:flucutation:types}
		\boldsymbol{\eta}_i(t) = \sqrt{2 d_E} \mbox{ $\left ( \! \! \!	\begin{array}{c}
 					  \xi_{i,x}(t)\\ 
 					  \xi_{i,y}(t)
				\end{array} \! \! \! \right ) $} + \sqrt{2 d_v} \, \xi_{i,v} (t) \, \bi{e}_{i,v}(t) + \sqrt{2 d_{\varphi}} \, \xi_{i,\varphi} (t) \, \bi{e}_{i,\varphi}(t) 	
	\end{eqnarray}
The first expression on the right hand side denotes so called passive or external noise acting in the same way on both components of the velocity vector $\bi{v}_i$ with the intensity $d_E$. It might be associated for example with ordinary Brownian fluctuations. Thus, in principle it can model essentially a equilibrium phenomenon linked to thermal fluctuations. In the present context, however, it may also be of non-equilibrium origin. We refer to this type of noise as passive, as it does not ``carry'' any informations on the propulsion of a particle. 

\begin{figure}[htb]
	\begin{center}
		\includegraphics[width=0.9\textwidth]{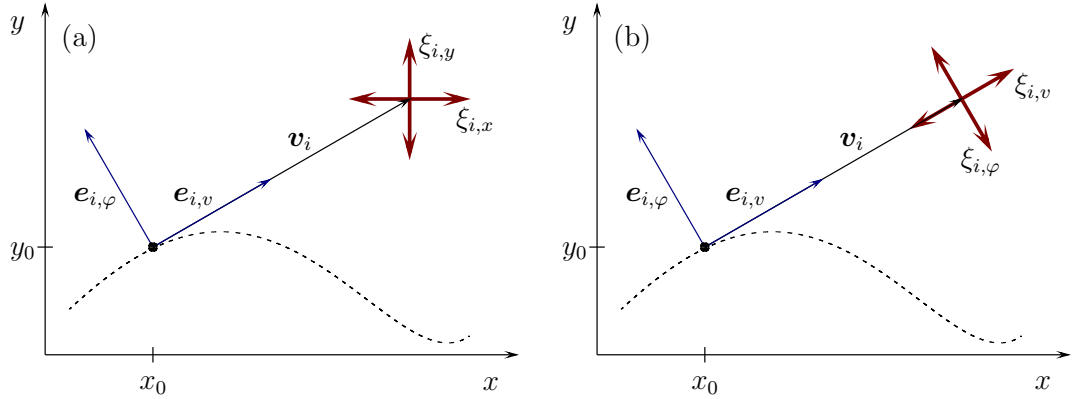}
		\caption{Visualization of passive (a) and active fluctuations (b). The dashed line shows the trajectory of the $i^{\mbox{\tiny{th}}}$ particles. Speed noise changes the speed of the particle only whereas angular noise acts on the direction of motion. On the other hand, the components of the velocity vector are changed by passive fluctuations in the same way.}
		\label{pic:ABM:fluctuaion:types}
	\end{center}
\end{figure}

We refer to the remaining two terms together as active or internal noise. It is motivated by the assumption, that this internal fluctuations are directly linked to fluctuations of the driving ``motor'' or internal decision processes of the Brownian agent, i.e. animals. These active fluctuations are subdivided in two parts: speed noise, acting along the direction of motion with intensity $d_v$ and pure angular fluctuations, changing the direction of motion of the particle with intensity $d_{\varphi}$ without affecting the particles speed. In the case of constant speed it is not possible to distinguish active and passive fluctuations as the angle defining the direction of motion is the only degree of freedom. 
%In the overdamped case of constant speeds ($\alpha \gg \mu$), the angular fluctuations are equivalent to the noise studied in the Vicsek model. 
Figure \ref{pic:ABM:fluctuaion:types} illustrates the different effects of active and passive fluctuations on the particles velocities. 

The random processes $\xi_{i,\nu}(t)$ are assumed to by Gaussian white noise with zero mean and vanishing correlations: $\mean{\xi_{i,\nu}(t)}=0$, $\mean{\xi_{i,\nu}(t)\xi_{j,\mu}(t')}=\delta_{ij} \delta_{\nu\mu} \delta \! \left (t-t'\right)$ ($i,j \! \in \! \left \{1,...,N \right \}$, $\nu,\mu \! \in \! \left \{x,y,v,\varphi \right \}$). 
Due to the non-equilibrium nature of the self-propelled dynamics and active fluctuations, the fluctuation-dissipation theorem does not hold here in general.

%Moreover, the velocity dynamics is highly nontrivial and not restricted to constant speeds. In absence of noise, all particle velocities are aligned in the long time limit and collective motion appears.  
	
\subsection{Natural units and dimensionless parameters}
\label{sec:natural:units}
First of all, we identify the relevant dimensionless parameters and characteristic length and time scales in \eqnref{eqn:ABM:active:particle:eqn}: 
%	\numparts		
		%\begin{eqnarray}
		\begin{equation}
			\mbox{L} = \left [\frac{m v_0}{\alpha} \right ], \qquad \mbox{T} = \left [\frac{m}{\alpha} \right] \, .
		\end{equation}
		%\end{eqnarray}
%	\endnumparts
Apparently, the velocity scale is defined by $v_0$. In this section, physical values in SI units are indicated by a prime. In order to reduce the number of parameters, the following transformation to dimensionless quantities ${\bi r}$, ${\bi v}$ and $t$ is used: 
	%\numparts
		\begin{eqnarray}
			\bi{r}' = \frac{m v_0}{\alpha} \, \bi{r} \qquad  
			t'	= \frac{m}{\alpha} \, t  \qquad 
			\bi{v}' = v_0 \, \bi{v} 
 		\end{eqnarray}
	%\endnumparts
The Gaussian processes $\xi(t')$ are transformed in the same manner: 
	\begin{eqnarray}
		\xi(t')=\sqrt{\frac{\alpha}{m}} \, \xi(t) \ .
	\end{eqnarray}
Using these transformations, one can rewrite the microscopic Langevin dynamics \eqnref{eqn:ABM:active:particle:eqn} of the $i^{\mbox{\footnotesize{th}}}$ particle in dimensionless quantities as follows: 
	\eqnlabel{eqn:KT:NU:eqn4}
	\numparts
		\begin{eqnarray}
			\fl \frac{\rmd \bi{r}_i}{\rmd t}  &=& \bi{v}_i \,  , 								\label{eqn:KT:NU:eqn4a} \\
			\fl \frac{\rmd \bi{v}_i}{\rmd t}  &=& \bi{e}_{i,v} - \bi{v}_i + \mu \left (\bi{u}_{A_i} - \bi{v}_i \right ) + \sqrt{2 D_E} \! \mbox{ $\left ( \! \! \!	\begin{array}{c}
 					  \xi_{i,x}\\ 
 					  \xi_{i,y}
				\end{array} \! \! \! \right ) $} + \sqrt{2 D_v} \, \xi_{i,v} \, \bi{e}_{i,v}  + \sqrt{2 D_{\varphi}} \, \xi_{i,\varphi} \, \bi{e}_{i,\varphi} \, . 	\label{eqn:KT:NU:eqn4b} 
		\end{eqnarray}
	\endnumparts
In \eqnref{eqn:KT:NU:eqn4}, the dimensionless velocity-alignment strength $\mu$  and noise intensities $D_\nu$ were introduced ($\nu \in \left \{E,v,\varphi \right \}$):
	%\numparts
		\begin{eqnarray}
			\mu &= \frac{\tilde{\mu}}{\alpha}\ , \qquad D_{\nu} &= \frac{d_{\nu}}{m \alpha v_0^2} .
		\end{eqnarray}
	%\endnumparts
Hence, the dynamical behaviour is governed by these four dimensionless parameters. 

\section{Hydrodynamic theory of active Brownian particles}
\label{sec:KT:ABM}
A hydrodynamic description based on symmetry properties of flocks was first proposed by Toner and Tu (TT) \cite{toner_long-range_1995,toner_flocks_1998}. They proposed a general continuum model for collective motion without considering a particular microscopic dynamics and determined the scaling exponents for density and velocity fluctuations. Hydrodynamic equations for SPP with aligning collisions were derived by Bertin \etal in \cite{bertin_boltzmann_2006,bertin_microscopic_2009} using a Boltzmann approach, which allows to link microscopic model parameters to the parameters of a coarse-grained theory of TT-type. Bertin \etal assumed a small collective velocity, which implies the validity of the theory only close to the transition point.  Recently, Ihle derived a kinematic description of the Vicsek model accounting also for multiparticle interactions \cite{ihle_kinetic_2011}. Related hydrodynamic descriptions of self-propelled rods were also studied by Marchetti and coworkers \cite{baskaran_hydrodynamics_2008,mishra_fluctuations_2010}.

In order to predict the macroscopic behaviour of the active Brownian particle gas, a hydrodynamic description is derived directly from the microscopic Langevin dynamics in the following. We use a moment expansion of the corresponding probability density function (see also \ref{sec:app:derivation_of_moment_eqn} for further details).
The approach used in this work goes back to Riethm\"uller \etal, who studied gravity-driven granular flow \cite{riethmueller_langevin_1997}. It was successfully used to analyze the mean-field dynamics of ABP with velocity alignment in one dimension \cite{romanczuk_collective_2010} and in two dimensions with passive fluctuations only \cite{romanczuk_mean-field_2011}. 

%\subsection{Deduction of hydrodynamic equations from the microscopic dynamics}
Let $p\left (\bi{r}_j,\bi{v}_j,t \right )\rmd^2 \bi{r}_j \rmd^2 \bi{v}_j$ be the probability to find the $j^{\mbox{\footnotesize{th}}}$ particle at time $t$ in the velocity interval $\left [\bi{v}_j,\bi{v}_j+\rmd \bi{v}_j \right ]$ and within the spatial region $\left [\bi{r}_j,\bi{r}_j + \rmd \bi{r}_j \right ]$. The dynamics of the one-particle probability density function $p\left (\bi{r}_j,\bi{v}_j,t \right )$ can easily be derived from the dynamics of the joint probability density function  $P = P \left (\bi{r}_1,\bi{r}_2,...,\bi{r}_N,\bi{v}_1,\bi{v}_2,...,\bi{v}_N,t \right)$ by integrating over all particle positions and velocities except the $j^{\mbox{\footnotesize{th}}}$.
Due to the fact that any $j$ could be chosen, the particle index is omitted henceforward. The dynamics of the single-particle probability density function $p\left (\bi{r},\bi{v},t\right )$ reads
	\begin{eqnarray}
		\label{eqn:KT:espFPE}
		 \fl \frac{\partial p}{\partial t} =& - \nabla_{\bi{r}} \left ( \bi{v} \, p \right ) - \nabla_{\bi{v}} \left [\left (\frac{\bi{v}}{\abs{\bi{v}}} - \bi{v}\right ) p + \mu \left (\bi{u}_{A} - \bi{v} \right ) p - D_{\varphi} \left (\frac{\bi{v}}{\; \abs{\bi{v}}^2} \right) p \right ] \nonumber  \\
		 \fl & + \, \frac{\partial^2}{\partial v_{x}^2} \left ( \frac{D_v v_{x}^2 + D_{\varphi} v_{y}^2}{\abs{\bi{v}}^2} \, p \right ) + \frac{\partial^2}{\partial v_{y}^2} \left ( \frac{D_v v_{y}^2 + D_{\varphi} v_{x}^2}{\abs{\bi{v}}^2} \,  p \right )  \nonumber \\
		 \fl & + \, \frac{\partial^2}{\partial v_{x} \partial v_{y}} \left [ 2 \,  \frac{v_{x} v_{y}}{\abs{\bi{v}}^2} \left ( D_v - D_{\varphi} \right ) p  \right ] + \Delta_{\bi{v}} \left ( D_{E} \, p \right ) \, \mbox{. }
	\end{eqnarray}	
Moments of the velocity components are defined by
\begin{equation}
\label{eqn:KT:moment:definition}
M^{(mn)} = \mean{v_x^m v_y^n} = \frac{1}{\rho(\bi{r},t)} \int \rmd^2 \bi{v} \; v_x^m v_y^n \; p(\bi{r},\bi{v},t)  \, \mbox{. }
\end{equation}
The spatial density of particles corresponds directly to $M^{(00)}$:
	\begin{eqnarray}
		\label{eqn:KT:local density}
		\rho (\bi{r},t) = \int \rmd^2 \bi{v} \; p(\bi{r},\bi{v},t) \,\mbox{. } 
	\end{eqnarray}
Multiplying \eref{eqn:KT:moment:definition} with the density and taking the temporal derivative yields
	\begin{eqnarray}
		\label{eqn:KT:ansatz:moment:expansion}
		\frac{\partial }{\partial t} \left ( \rho \, M^{(mn)} \right ) = \int \rmd^2 \bi{v} \; v_x^m v_y^n \; \frac{\partial p}{\partial t}  \, \mbox{. }
	\end{eqnarray}
We may rewrite the integral on the right-hand side of \eref{eqn:KT:ansatz:moment:expansion} by inserting the Fokker-Planck equation \eref{eqn:KT:espFPE} and using the definition (\ref{eqn:KT:moment:definition}) to obtain a general equation of motion for the $M^{(mn)}$-moment of the probability density function: 
	\begin{eqnarray}
		\label{eqn:KT:moment:equation}
		\fl \frac{\partial }{\partial t} \left ( \rho \mean{v_x^m v_y^n} \right ) = 
		    & - \frac{\partial}{\partial x} \left (\rho \mean{v_x^{m+1}v_y^n} \right ) - \frac{\partial }{\partial y} \left( \rho \mean{v_x^m v_{y}^{n+1}} \right) 
		\nonumber \\
		\fl & + \, (m+n) \rho \left[ \mean{\frac{v_x^m v_y^n}{\abs{\bi{v}}}} - \mean{v_x^m v_y^n} - D_\varphi \mean{\frac{v_x^m v_y^n}{\abs{\bi{v}}^2}} \right] 
		\nonumber \\
		\fl & + \mu \rho \left[ m\, u_{A,x}\mean{v_x^{m-1}v_y^n} + n\, u_{A,y}\mean{v_x^{m}v_y^{n-1}} - (m+n)\mean{v_x^m v_y^n} \right]  
		\nonumber \\
		\fl & + \,  m(m-1) \rho \left [D_{\varphi} \mean{\frac{v_x^{m-2} v_{y}^{n+2}}{\abs{\bi{v}}^2}} + D_v \mean{\frac{v_x^m v_y^n}{\,\abs{\bi{v}}^2}}  + D_E \mean{v_x^{m-2} v_y^{n}} \right ]
		\nonumber \\
		\fl & + \,  n(n-1) \rho \left [D_{\varphi} \mean{\frac{v_x^{m+2} v_{y}^{n-2}}{\abs{\bi{v}}^2}} + D_v \mean{\frac{v_x^m v_y^n}{\,\abs{\bi{v}}^2}}  + D_E \mean{v_x^{m}v_y^{n-2}} \right ]
		\nonumber \\
		\fl & +2 m n \rho \mean{\frac{v_x^{m} v_{y}^{n}}{\abs{\bi{v}}^2}}\left(D_v-D_\varphi \right) \, \mbox{.}
	\end{eqnarray}
Up to now no explicit approximations were made except the mean field assumption ($N_{A_i} \gg 1$).  	

In order to derive hydrodynamic equations from \eref{eqn:KT:moment:equation}, we rewrite the individual velocity $\bi{v}$ as a sum of a mean-field velocity $\bi{u} (\bi{r},t )$ at the position of the focal particle plus some deviation $\bdelta \bi{v} ( \bi{r},t )$:
	\begin{eqnarray}
		\bi{v} = \bi{u}(\bi{r},t) + \bdelta \bi{v} (\bi{r},t) \, \mbox{. }
	\end{eqnarray}
We assume that the expectation value of $\delta {v}^k_{x,y} ( \bi{r},t )$ vanishes for odd $k$:
	\begin{eqnarray}\label{eq:deltav}
		\mean{\delta {v}^k_{x,y}} = 0 \quad \mbox{for} \quad k=1,3,5,\dots \ .
	\end{eqnarray}
This corresponds to the assumption of a symmetric $p(\bi{r},\bi{v},t)$ with respect to $\bi{u} (\bi{r},t )$ in the velocity space.
Furthermore, we assume
	\begin{eqnarray}
		\mean{\left (\delta v_{x}\right)^2 \delta v_y} & = \mean{\left (\delta v_{y}\right)^2 \delta v_x} = 0 \, \mbox{ .}
	\end{eqnarray}	
In addition to the mean-field velocity, we introduce the covariance matrix as a measure of velocity fluctuations around the mean-field velocity:
\begin{eqnarray}
		\mbox{
			$\hat{\bi{T}}
			= \left (
				\begin{array}{cc}
 					\mean{\left (\delta v_x \right)^2} & \mean{\delta v_x \delta v_y}  \\ 
 					\mean{\delta v_y \delta v_x} & \mean{\left (\delta v_y \right)^2} 
				\end{array}
			 \right ) 
			= \left (
				\begin{array}{cc}
 					T_{xx} & T_{xy}  \\ 
 					T_{yx} & T_{yy} 
				\end{array}
 			  \right ) 
			$
		} = \hat{\bi{T}}^{\,T} \ .
	\end{eqnarray}
In the following we will refer to $\hat{\bi{T}}$ as the  ``temperature tensor''.

In the lowest order ($m=n=0$), one obtains from \eref{eqn:KT:moment:equation} the continuity equation for the local density  
	\begin{eqnarray}
		\label{eqn:KT:cons:of:N}
		\frac{\partial \rho}{\partial t} + \nabla_{\bi{r}} \left ( \rho \, \bi{u} \right ) = 0 \, \mbox{, }
	\end{eqnarray}
reflecting the conservation of the particle number $N$. 

%Averaging over the fractional terms involves an infinite number of moments for what reason an appropriate closure has to be found. In that case, numerator and denominator are averaged independently of each other. Furthermore it is assumed, that the order of extracting the root and averaging is commutating, i.e.  
%The denominator of the last term will be abbreviated in the following.
%	\begin{eqnarray}
%		V = \sqrt{\abs{\bi{u}}^2 + T}
%	\end{eqnarray}
%The hydrodynamic equations \eref{eqn:KT:cons:of:N}-\eref{eqn:KT:eqn:for:T} are rewritten appyling \eref{eqn:KT:approximation2}.

To obtain evolution equations for the mean-field velocity and the temperature tensor, we need  to evaluate the expectation values of fractional expressions of the form $v_x^m v_y^n/|{\bi v}|^k$ in \eref{eqn:KT:moment:equation}. Here, we use the approximation %\eref{eqn:KT:moment:equation}
\begin{eqnarray}
	\label{eqn:KT:approximation2}
	\mean{\frac{v_x^m v_y^n}{\; \abs{\bi{v}}^k}} \approx \frac{\mean{v_x^m v_y^n}}{\mean{\sqrt{v_x^2 + v_y^2}^{\, k}}} \approx \frac{\mean{v_x^m v_y^n}}{V^k} 
\end{eqnarray}
with
%\begin{equation}
$ V=V({\bi{u}},{\hat{\bi{T}}})=\sqrt{\abs{\bi{u}}^2 +  \mbox{tr}(\hat {\bi T})} \, .$
%\end{equation}
This approximation provides naturally a closure of the hierarchy of moment equations  with $m+n\leq2$. 

From equation (\ref{eqn:KT:ansatz:moment:expansion}) for the first order moments with $m = 1$, $n=0$ and $m=0$, $n=1$ combined with \eref{eqn:KT:cons:of:N}, we can derive the evolution equation for the mean-field velocity field to
	\begin{eqnarray}
	\label{eqn:KT:eqn:for:u}
  		\fl \frac{\partial \bi{u}}{\partial t} + \left ( \bi{u} \nabla_{\bi{r}} \right ) \bi{u} = 
			  \left ( \frac{1}{V}-1\right ) \bi{u} - \frac{D_\varphi}{V^2} \bi{u} + \mu \left (\bi{u}_{A} - \bi{u} \right ) 
			- \frac{1}{\rho}\left[ \nabla^T ( \rho {\hat {\bi T}}) \right]^T \, .
\end{eqnarray}
%\eref{eqn:KT:moment:equation}
The above equation \eref{eqn:KT:eqn:for:u} is a partial integro-differential equation since 
	\begin{eqnarray}
		\label{eqn:KT:integro:diff}
		\bi{u}_{A}(\bi{r},t) = A^{-1} \int_{A} \rmd^2 \bi{r}' \; \bi{u}(\bi{r}',t). 
	\end{eqnarray}
Csah{\'o}k and Czir{\'o}k \cite{csahok_hydrodynamics_1997} have suggested a hydrodynamic velocity equation for collective motion of bacteria with similar active propulsion and interaction terms as derived in Eq. \eref{eqn:KT:eqn:for:u}. We should note that in general, for a finite numbers of neighbours ${\bi u}_\epsilon$, will be a fluctuating quantity which is neglected here by assuming $N_\epsilon \gg 0$.
 
Finally, we can combine \eref{eqn:KT:cons:of:N} and \eref{eqn:KT:eqn:for:u} together with the three second-order equations obtained from \eref{eqn:KT:moment:equation} for $m+n=2$ ($ (m,n)\in \{(2,0),(0,2),(1,1)\}$). This yields  the equation for the temperature tensor
	\begin{eqnarray}
		\label{eqn:KT:eqn:for:T}
		\fl \frac{1}{2} \left [ \frac{\partial {\hat {\bi T}}}{\partial t} + \left ( \bi{u} \nabla_{\bi{r}}\right ) \hat {\bi T}  \right ] = &
		\left( \frac{1}{V} - 1\right){\hat {\bi T}}  - \mu {\hat {\bi T}} + D_E {\hat {\bi I}} + \frac{D_v}{V^2}\hat {\bi B}_v - \frac{D_\varphi}{V^2} {\hat {\bi B}}_\varphi  
	\nonumber \\
	& +\frac{1}{2}\left[ (\hat {\bi T} \nabla){\bi u}^T  + \left( ({\hat {\bi T}}\nabla){\bi u}^T\right)^T \right]\ ,  
	\end{eqnarray}
with $\hat {\bi I}$ being the identity matrix and 
	\numparts
	\begin{eqnarray}
		\hat {\bi B}_v  =  &({\bi u}{\bi u}^T+{\hat {\bi T}}) = \left( \begin{array}{cc} u_x^2+T_{xx} &  u_xu_y+T_{xy} \\ u_xu_y+T_{xy} & u_y^2+T_{yy}\end{array}\right), \\
		\hat {\bi B}_\varphi = & \left( \begin{array}{cc} u_y^2+T_{yy}-T_{xx} &  u_xu_y-2 T_{xy} \\ u_xu_y-2 T_{xy} & u_x^2+T_{xx}-T_{yy}\end{array}\right). 	
	\end{eqnarray} 
	\endnumparts
From \eref{eqn:KT:eqn:for:u} and \eref{eqn:KT:eqn:for:T} one can already identify the general impact of the different fluctuation types: The active angular fluctuations $D_\varphi$ introduce an additional effective friction which affects both the velocity and temperature dynamics, whereas  the active velocity fluctuations $D_v$ and passive fluctuations $D_E$ drive the temperature equation.

The kinetic theory contains the important limiting case of ordinary Brownian motion, which corresponds to $\mu$, $D_{\varphi}$, $D_v$ $v_0$ equal to zero. In this special case, we recover the fluctuation-dissipation relation known from equilibrium statistical mechanics.

\section{Spatially homogeneous, steady-state solutions}

Under the assumption of a spatially homogeneous system, we may drop all spatial derivatives in \eref{eqn:KT:cons:of:N}, \eref{eqn:KT:eqn:for:u} and \eref{eqn:KT:eqn:for:T}. 
Furthermore, in the steady-state case, we may choose the coordinate system such that $\bi{u}=\left (\abs{\bi{u}} \!,0 \right)$ is parallel to the $x$-axis of the laboratory frame of reference. In this case, we assume that the average deviations parallel and perpendicular to the direction of motion become independent $\mean{\delta v_x \delta v_y} = \mean{\delta v_\|  \delta v_\bot}=  \mean{\delta v_\|} \mean{\delta v_\bot}$, i.e. the temperature tensor assumes a diagonal form
	\begin{eqnarray}
		\mbox{
			$\hat{\bi{T}}
			= \left (
				\begin{array}{cc}
 					\mean{(\delta v_\|)^2} & \mean{\delta v_\|}\mean{\delta v_\bot}  \\ 
 					\mean{\delta v_\bot}\mean{\delta v_\|} & \mean{(\delta v_\bot)^2} 
				\end{array}
			 \right ) 
			= \left (
				\begin{array}{cc}
 					\Tp & 0  \\ 
 					0 & \Ts 
				\end{array}
 			  \right ) \, \mbox{, } 
			$ 
		}
	\end{eqnarray}
where $\Tp$ is the mean squared velocity deviation parallel to $\bi{u}$, whereas $\Ts$ is the mean squared deviation perpendicular to $\bi{u}$. We define $T=\tr \, ( \hat{\bi{T}})$ as the (scalar) kinetic temperature. 
The simplifying assumption of independent average fluctuations parallel and perpendicular to the average velocity is justified by the broken symmetry of the collective translational steady-state \cite{toner_flocks_1998}. Static correlations in the macroscopic mean-field velocity fluctuations would imply collective rotational motion (see e.g. \cite{romanczuk_brownian_2011}), which are not stable in our model.

In the spatially homogeneous and stationary case, we obtain the following set of homogeneous equations. 
	\numparts
		\begin{eqnarray}
		 \fl	\frac{\rmd \!\abs{\bi{u}}}{\rmd t}  &= 0 =&  \left (\left (\abs{\bi{u}}^2 + T\right)^{-\frac{1}{2}} - 1 - D_{\varphi} \left (\abs{\bi{u}}^2 + T\right)^{-1} \right) \abs{\bi{u}} \label{eqn:KT:u:eqn} \\
		 \fl	\frac{1}{2} \frac{\rmd \Tp}{\rmd t} &= 0 =& \left ( \left (\abs{\bi{u}}^2 + T\right)^{-\frac{1}{2}} - 1 - \mu \right)\Tp + D_{\varphi} \left ( \frac{\Ts - \Tp}{\abs{\bi{u}}^2 + T} \right )  \nonumber \\
		 \fl &&  
+ D_v \left ( \frac{\abs{\bi{u}}^2 + \Tp}{\abs{\bi{u}}^2 + T} \right) + D_E \label{eqn:KT:tp:eqn}\\
		 \fl	\frac{1}{2} \frac{\rmd \Ts}{\rmd t} &= 0 =& \left ( \left (\abs{\bi{u}}^2 + T\right)^{-\frac{1}{2}} - 1 - \mu \right)\Ts + D_{\varphi} \left ( \frac{\abs{\bi{u}}^2 +\Tp - \Ts}{\abs{\bi{u}}^2 + T} \right ) \nonumber  \\
		 \fl && 
+ D_v \left ( \frac{\Ts}{\abs{\bi{u}}^2 + T} \right)  + D_E \label{eqn:KT:ts:eqn} \ .
		\end{eqnarray}
	\endnumparts
The corresponding stationary solutions of the ordinary differential equations \eref{eqn:KT:u:eqn}-\eref{eqn:KT:ts:eqn} can in principle be found but are quite complex and hard to analyze. A simpler equation for the temperature can be found by adding up \eref{eqn:KT:tp:eqn} and \eref{eqn:KT:ts:eqn}, which yields 
	\eqnlabel{eqn:KT:global:mean:field:eqn}	
	\numparts
		\begin{eqnarray}
			\fl \frac{\rmd \!\abs{\bi{u}}}{\rmd t}  &= 0 =&  \left (\left (\abs{\bi{u}}^2 + T\right)^{-\frac{1}{2}} - 1 - D_{\varphi} \left (\abs{\bi{u}}^2 + T\right)^{-1} \right) \abs{\bi{u}}\ , \\
			\fl \frac{1}{2} \frac{\rmd T}{\rmd t} &= 0 =& \left (\left (\abs{\bi{u}}^2 + T\right)^{-\frac{1}{2}} - 1 - \mu \right) T  + D_{\varphi} \abs{\bi{u}}^2 \left(\abs{\bi{u}}^2 + T \right)^{-1} + D_v + 2 D_E \ .
		\end{eqnarray}
	\endnumparts
%The many-body problem is described by two macroscopic scalar variables: the mean speed $|\bi{u}|$ (order parameter) and the temperature $T$ (magnitude of fluctuations). 

In the following, we discuss the time-independent stationary solutions of \eqnref{eqn:KT:global:mean:field:eqn} which correspond to fixed points of the dynamical system. One pair of fixed points reads
	\eqnlabel{eqn:KT:disordered:state}
	\numparts
		\begin{eqnarray}
			\fl \abs{\bi{u}}_1 &= \, 0 \, \mbox{, } \\
			\fl T_1^{\,\left (\pm \right)} &= \, \frac{2 D_E + D_v}{1 + \mu} + \frac{1}{2} \left ( 1+\mu \right )^{-2} \left ( 1 \pm \sqrt{1 + 4 \, \left (2 D_E + D_v \right ) \left (1 + \mu \right)} \; \right ) \, \mbox{. }
		\end{eqnarray}
	\endnumparts
However, only one of these fixed points is physically relevant. Solution $\left (\abs{\bi{u}}_1,  T_1^{\,\left (- \right)} \right)$ will be neglected because it predicts unphysical behaviour for vanishing noise intensities. 
In this case, both $\abs{\bi{u}}_1$ and $T_1^{\,\left (- \right)}$ are equal to zero. This corresponds to all particles being at rest, which is not possible for permanently self-driven particles considered here. 
Solution $\left (\abs{\bi{u}}_1,  T_1^{\,\left (+ \right)} \right)$ with vanishing mean speed and finite fluctuations is called disordered state. It is stable above a critical noise intensity and below a critical coupling strength, respectively. The temperature in the disordered state is independent of the intensity of angular fluctuations $D_{\varphi}$. 

Another two solutions with nonzero mean speed are given by the following expressions: 
	\eqnlabel{eqn:KT:ordered:state}
	\numparts
		\begin{eqnarray}
			\abs{\bi{u}}_2^{\left ( \pm \right )} &= \, \sqrt{ \frac{1}{2} \left (1 \pm \sqrt{1 - 4 D_{\varphi}} \, \right ) - D_{\varphi} - \frac{D_{\varphi} + D_v + 2 D_E}{\mu}} \, , \\
			T_2 &= \, \frac{D_{\varphi} + D_v + 2 D_E}{\mu} \, .
		\end{eqnarray}
	\endnumparts
We refer to these fixed points as ordered state. For vanishing noise intensities ($D_E=D_v=D_\varphi=0$), one solution reads $u_2^{\left (+\right)}=1$ and $T_2 = 0$. It corresponds to all particles moving perfectly aligned in the same direction with stationary speed of individual particles $\abs{\bi{v}_i}=1$. In general, the solution $u_2^{(+)}$ is stable for sufficiently weak noise, or strong alignment, respectively. The second ordered solution $u_2^{(-)}$ is always unstable.  

	\begin{figure}%[H]
		\begin{center}
			\includegraphics[width=0.95\textwidth]{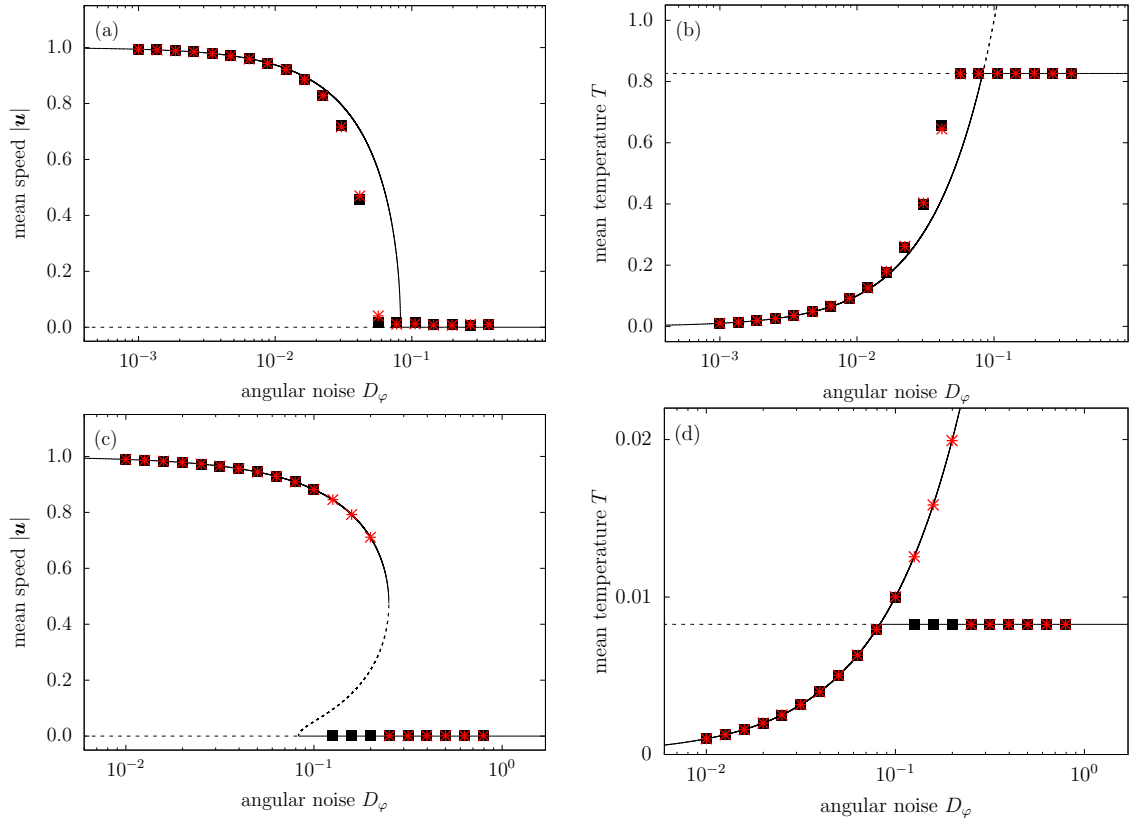}
			\caption{Comparison of numerical Langevin simulations of \eqnref{eqn:KT:NU:eqn4} (points) and theoretical prediction \eqnref{eqn:KT:global:mean:field:eqn} (lines) in the case of vanishing speed noise and passive noise ($D_E = D_v = 0$). The dashed lines indicates unstable states whereas the solid lines represent stable configurations. Results of simulations initially started in the disordered state are indicated by (black) squares. Stars (red) represent results of simulations which were performed using the ordered state as initial condition. The simulations were performed considering global interaction $A_i = L\times L$, periodic boundary conditions and $N=10^4$ particles. Parameter: $\mu = 0.1$ (a)-(b), $\mu = 10$ (c)-(d), numerical time step $\Delta t=1\cdot10^{-4}$. }
			\label{pic:KT:d_phi_only:sim:res}
		\end{center}
	\end{figure}

In \eref{eqn:KT:global:mean:field:eqn}, the intensities of passive noise $D_E$ and active speed noise $D_v$ enter the equation with the same sign. The theory predicts a qualitatively identical influence of passive fluctuations and speed noise on the mean-field behaviour of the system. 

\begin{figure}%[htb]
		\begin{center}
			\includegraphics[width=0.8\textwidth]{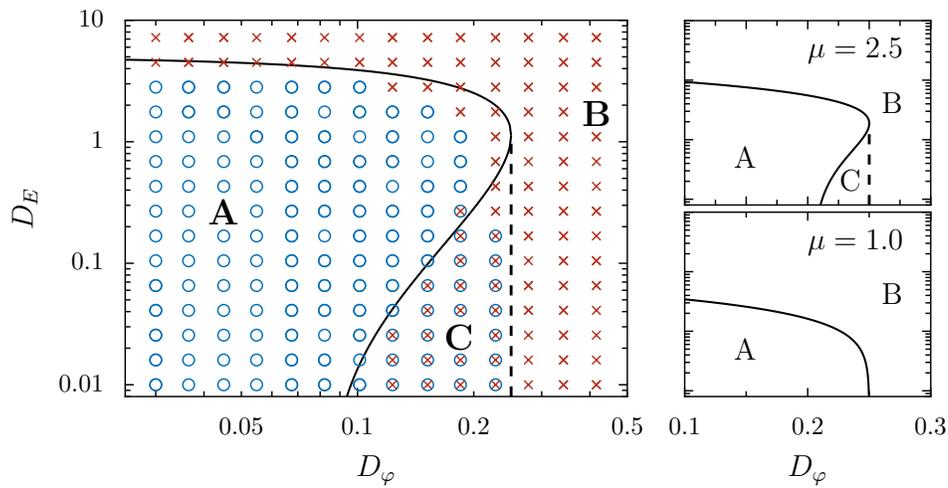}
			\caption{Phase diagram in the case of global interaction for $D_v=0$. Three different regimes can be distinguished: only ordered solution stable (A), only disordered solution stable (B), and a bistable regime (C). The bistable regime, where both the ordered and disordered states are stable, exists only for finite active angular noise intensities and strong interaction $\mu > 1$, e.g. for $D_E=0$: $\mu(\mu + 1)^{-2} < D_{\varphi} < 0.25$.  It does not exist for the case of passive fluctuations only \cite{romanczuk_mean-field_2011}. The solid lines represent the borders of the different regimes as predicted by the theory. The symbols represent the results of numerical simulations with $N=10^4$ particles. At each point in the phase space two simulation runs were performed with different initial conditions (ordered and disordered) until the system reached a stationary state. The circles represent the ordered stationary result, whereas the crosses show the results with a disordered final state. The dashed line shows the critical active angular noise $D_{\varphi,c}=0.25$. Numerical time step in all simulations $\Delta t=5\cdot10^{-3}$.  
}
			\label{pic:KT:phase_diagram}
		\end{center}
	\end{figure}

For weak interaction $\mu<1$, the theory always predicts a continuous transition of the mean velocity from the ordered to the disordered state (figure \ref{pic:KT:d_phi_only:sim:res}a,b). 
In the case of strong interaction $\mu > 1$ and not too large passive fluctuations, a new regime emerges. In this regime, both the ordered and the disordered state are stable macroscopic configurations (figure \ref{pic:KT:d_phi_only:sim:res}c,d). The theory predicts that, depending on the initial condition, the system relaxes into one of these stable macroscopic states. Therefore, a discontinuous transition from collective motion to the disordered state for increasing angular fluctuations is expected.
In figure \ref{pic:KT:phase_diagram}, we show the phase diagram with respect to external noise $D_E$ and active angular fluctuations $D_\varphi$ for $\mu=10$ and $D_v=0$. 
In the bistable regime, if the mean velocity $\bi{u}_{A}={\bi u}$ is low, the velocity-alignment force reduces to Stokes friction (social friction) which dominates the individual particle dynamics and as a consequence leads to a stable disordered state.
Surprisingly, the transition point $D_{\varphi,c} = 0.25$ does not depend on the coupling strength $\mu$. From the Fokker-Planck equation \eref{eqn:KT:espFPE} one can see, that $D_\varphi>0$ leads to an effective drift against the collective direction of motion of the particles. Subsequently, the individual particle velocities decrease and above $D_{\varphi,c}$ no self-driven motion along the consensus direction motion is possible. 
%Slow moving particles are able to change their direction of motion faster, which in turn destroys the velocity correlations between different particles.  

Please note that the mean speed in the limiting case of strong interaction $\mu \rightarrow \infty$ still depends on the intensity of the angular fluctuations. 
	\begin{eqnarray}
		\lim_{{\mu \rightarrow \infty}} \abs{\bi{u}}_2^{\left ( \pm \right )} = \sqrt{ \frac{1}{2} \left (1 \pm \sqrt{1 - 4 D_{\varphi}} \, \right ) - D_{\varphi}}
	\end{eqnarray}

The analytical results \eqnref{eqn:KT:disordered:state} and \eqnref{eqn:KT:ordered:state} are compared with numerical simulations of the Langevin dynamics \eqnref{eqn:KT:NU:eqn4}. 
We used the stochastic Heun-Scheme for the numerical intergration, which converges in quadratic mean to the solution of Langevin equations interpreted in the sense of Stratonovich \cite{ruemelin_numerical_1982}. 
The simulations were performed with $N=10^4$ particles considering a square spatial area of length $L$ and periodic boundary conditions. To ensure a spatially homogeneous system, 
global interaction was used ($A_{i} = L\times L$).  In order to detect the bistable region, both the ordered and disordered state were used as initial condition. Results of the numerical simulations in comparison with analytical predictions of the mean-field theory are shown in figures \ref{pic:KT:d_phi_only:sim:res} and \ref{pic:KT:phase_diagram}. 

\begin{figure}
\begin{center}
\includegraphics[width=0.7\textwidth]{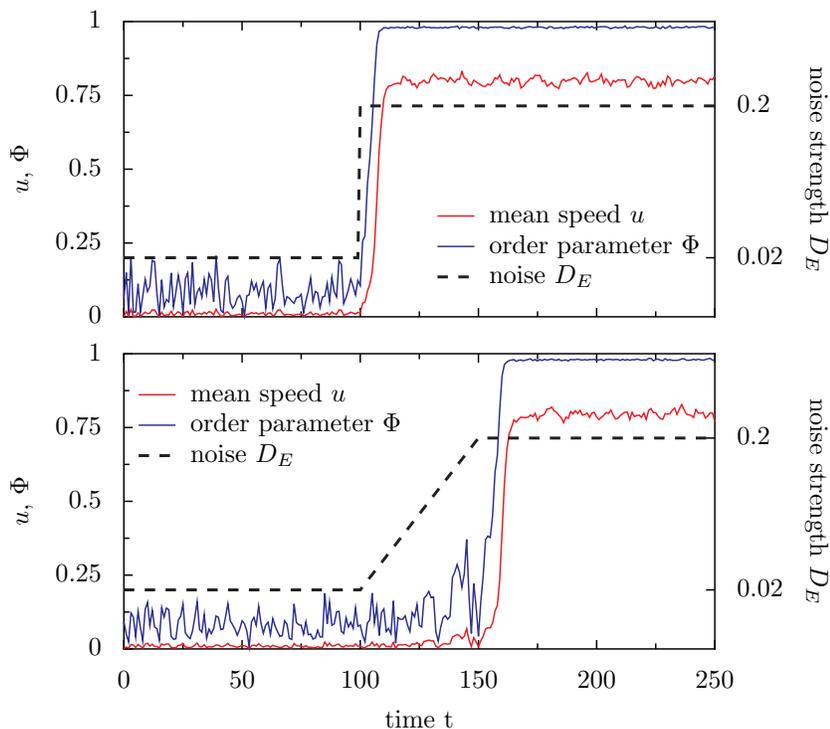}
\caption{Switching of the globally coupled system from the disordered state to the ordered state with increasing noise intensity for step like noise change (top) and a linear switching protocol (bottom). The system is initialized in the disordered state with a noise intensity of the passive fluctuations $D_E=0.02$ corresponding to the bistable regime. At time $t=100$, the noise intensity is increased either due to an instantaneous step-like switch or in finite time via a linear protocol to the final value $D_E=0.2$ which corresponds to the ordered regime. In addition to the mean speed $|{\bi u}|=u$, we show also the directional order parameter $\Phi=\left|\sum_{i=1}^N {\bi e}_{i,v}\right|/N$ with ${\bi e}_{i,v}={\bi v}_i/|{\bi v}_i|$. Other parameters: $D_\varphi=0.2$, $D_v=0$, $N=10^4$, $\mu=10$.}
\label{fig:dswitch}
\end{center}
\end{figure}

In the absence of active fluctuations, the theory is mainly confirmed by the numerical simulations. This case was discussed in detail in  \cite{romanczuk_mean-field_2011}. If the intensity of the (active) speed noise is small in comparison to intensities of the active angular and passive fluctuations, the theoretical predictions are in good agreement with the numerical results.

Deviations from the theory are due to the approximations used, especially \eref{eqn:KT:approximation2}. This approximation implies that the dynamics of the local velocity field follows closely the individual dynamics or more specifically: the decrease of the mean speed is related to a decrease of the individual particles velocities. Thus, the agreement of theory and simulations is better for a strong velocity-alignment, whereas small deviations are found in the case of fixed particles speeds ($\mu \ll 1$). 

According to the phase diagram in figure \ref{pic:KT:phase_diagram}, it is possible to move from the bistable regime (B) into the only ordered regime by increasing the external noise strength $D_E$. This implies the rather counter-intuitive effect of ``generating'' order from disorder by increasing noise. If we consider the system to be initially in the disordered state in the bistable regime, an increase in $D_E$ will eventually lead to a destabilization of the current state and to a transition from disordered to ordered solution. The actual transition depends on the protocol $D_E(t)$ as shown in figure \ref{fig:dswitch}. However, if the noise increase is not to strong, the system will eventually relax to the ordered state of collective motion.  

We point out that the found bistable behaviour differs from the
previously reported bistabilities in collective motion (see e.g. \cite{romanczuk_collective_2010,strefler_swarming_2008}), which were either a consequence of a non-linear friction function in one-dimensional dynamics \cite{romanczuk_collective_2010} or were caused by Morse-type interactions of active particles in three dimensions \cite{strefler_swarming_2008}. 
Noise in the cited cases did not create the coexistence of the two phases but merely induced switching between the two states with switching times which appeared to be exponentially distributed.

\section{Discussion \& Outlook}

%In this work, we consider a system of active Brownian particles with velocity alignment, where individual particles are subject to different noise types. 
%In particular, we distinguish two generic types of stochastic forces based on their correlation with the direction of motion of the individual particles: 
%1) ``active'' fluctuations which consist of independent components parallel and perpendicular to the particle heading, and 2) ``passive'' fluctuations 
%which are entirely independent on the direction of motion of the individual.
%
%We derive from the microscopic dynamics a hydrodynamic theory for the coarse-grained observables: spatial density $\rho$, velocity-field ${\bi u}$ 
%and the temperature tensor ${\hat {\bi T}}$, in the presence of both fluctuation types. 
%Our approach is based on the moment expansion of the single-particle probability density function solving the nonlinear Fokker-Planck equation 
%corresponding to the microscopic Langevin dynamics \cite{romanczuk_collective_2010,romanczuk_mean-field_2011}.      
%
In this work, we focused on the impact of the different fluctuation types on the collective dynamics of active Brownian particles with velocity alignment. We derived a hydrodynamic theory for the coarse-grained observables spatial density $\rho$, velocity-field ${\bi u}$ 
and the temperature tensor ${\hat {\bi T}}$ in the presence of both fluctuation types.
%We show that is is possible to link macroscopic variables to microscopic model parameters by the derivation hydrodynamic equations systematically from the Langevin dynamics. 
However, approximations are necessary in order to obtain a closed set of equations, which may lead to deviations of theoretical predictions from the numerical simulations.
For simplicity, we restricted ourselves to stationary, spatially homogeneous systems where analytical results can be obtained. This case corresponds either to global coupling (interaction length comparable to the system size) or to local dynamics of the system at length-scales where gradients in the macroscopic variables can be neglected. 
Our analysis shows that the introduction of active fluctuations has a striking influence on 
the mean-field behaviour of active particles with velocity-alignment and variable speed: In the case of strong interaction $\mu > 1$, 
a new bistable regime emerges that cannot be found for passive noise only. The corresponding transition from the ordered to the disordered state is discontinuous for increasing angular noise, whereas it is always continuous in the mean-field for purely passive noise \cite{romanczuk_mean-field_2011}. 
Our analytical results agree well with the simulation results for strong alignment $\mu\gg1$ and low active speed fluctuations $D_v$ which is related to the approximation used for the expectation values of rational expressions appearing in the general moment equation and the corresponding implicit moment closure.

The onset of order in the spatially homogeneous system is related to the synchronization of coupled oscillators \cite{birnir_ode_2007,paley_oscillator_2007}. 
If we identify the modulus of the velocity with the amplitude and the direction of motion with the phase of an oscillator, the active particle system with velocity-alignment corresponds to a 
system of coupled noisy oscillators with the same eigenfreqency in a co-rotating frame of reference. 
In the limit of small coupling $\mu\ll1$, the critical noise strengths for the transition from order to disorder have to be small as well ($D_E,D_v,D_\varphi \ll1$). 
Thus, the fluctuation in the modulus of the velocity at the critical point will be negligible. The system reduces effectively to coupled identical phase-oscillators which is a special case of the well studied Kuramoto-model \cite{kuramoto_self-entrainment_1975}. In this context, our results for strong alignment can be linked to amplitude death in system of strongly coupled limit-cycle oscillators \cite{mirollo_amplitude_1990}. The effective social friction in the disordered state due to the velocity alignment slows down the individual particles and stabilizes the disordered regime.

We focused here on the case of global interactions, however the hydrodynamic theory provides the basis of the stability analysis of the spatial inhomogeneous solution. 
It is known from the Vicsek model that density fluctuations occur in the case of local velocity-alignment interaction \cite{gregoire_onset_2004}. Recently, a number of studies analyzed the coarse-grained dynamics of Vicsek-type models showing the instability of the ordered homogeneous state of collective motion close to the transition point \cite{bertin_boltzmann_2006,bertin_microscopic_2009,lee_fluctuation-induced_2010,ihle_kinetic_2011}. 
Similar density inhomogeneities can be observed for active Brownian particles with local coupling and passive noise for sufficiently small interaction regions ($A_{i} \ll L\times L$) \cite{romanczuk_mean-field_2011}. 
Although our results for the homogeneous system in the presence of active fluctuations cannot be directly generalized for the case of local interactions, they are nevertheless relevant for it, as they provide predictions on local dynamics in high density regions. In the regime where only the ordered solution is stable (e.g. for purely passive fluctuations), regions of high density are correlated with regions of high order. 
Thus, they correspond to clusters or bands performing collective translational motion. This is also the case for the classical Vicsek model \cite{gregoire_onset_2004}. However, the situation changes drastically in the bistable regime. Here, high density structures may be both ordered (mobile) or disordered and stationary. 

\begin{figure}
\begin{center}
\includegraphics[width=\textwidth]{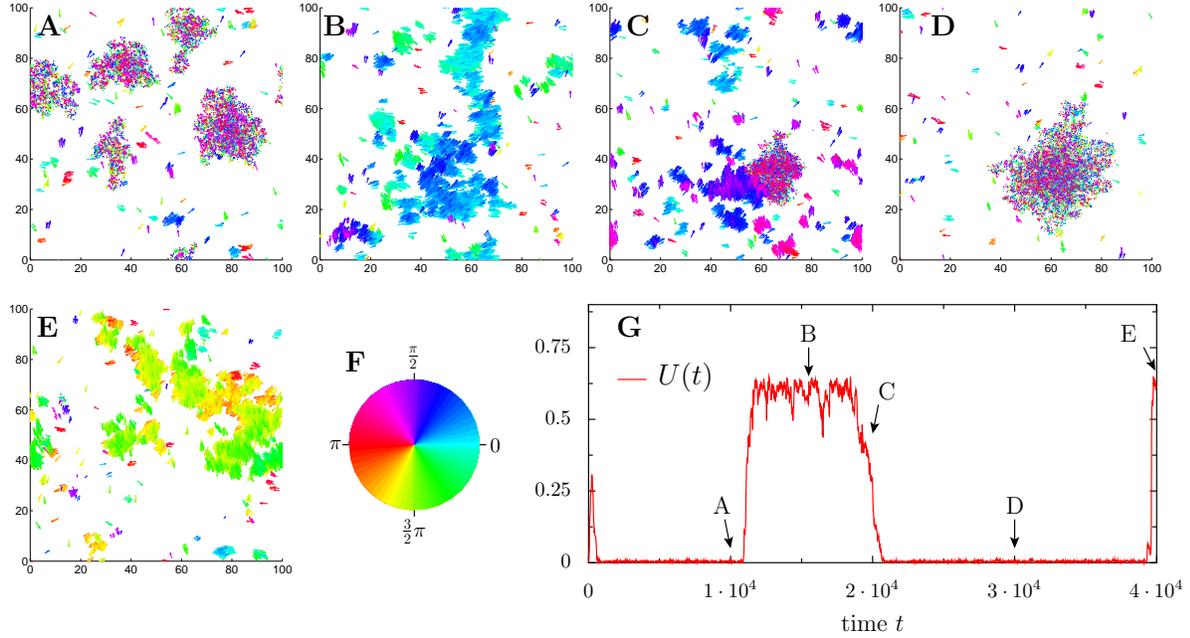}
\caption{Active Brownian particles with local coupling (disc-like interaction zone $A_i$ with interaction radius $\epsilon=1$, $L=100$, $N=10000$) in the bistable regime. (A-E): Spatial snapshots at $t=10000$ (A), $t=15500$ (B), $t=20000$ (C), $t=30000$ (D) and $t=39900$ (E). Each particle is shown by its velocity vector, with the color indicating the direction of motion according to (F). (G): Time evolution of the total mean velocity $U(t)=\left|\sum_i=1^N {\bi v}_i \right|$ of the corresponding simulation run with arrows indicating the snapshot times. The initial condition was a random distribution of particles and their directions of motion (disordered state). Other parameters: $D_\varphi=0.2$, $D_v = 0.0$, $D_E=0.01$, $\mu=10$, numerical time step $\Delta t=5\cdot10^{-3}$.
\label{fig:snapshots_local}}
\end{center}
\end{figure}

In figure \ref{fig:snapshots_local}, we show an example of a simulation for local coupling. We assume that a focal particle interacts with its neighbours within a metric range $\epsilon=1$ ($L=100$). The parameters are chosen to correspond to the mean-field bistable regime ($\mu=10$, $D_\varphi=0.2$, $D_v=0.0$, $D_E=0.01$). After a short initial transient, the system reaches a metastable state of multiple stationary clusters at $t\approx1000$, with few small mobile clusters in between (figure \ref{fig:snapshots_local}A). At time $t\approx11000$ a sufficiently large mobile cluster forms due to fluctuations. It is able to ``break up'' the stationary high density aggregates and we observe a sudden onset of large scale collective motion (figure \ref{fig:snapshots_local}A). At time $t\approx20000$, we observe again a spontaneous formation of a stationary (disordered) cluster due to a collision of clusters moving locally in different directions at $x\approx70$, $y\approx30$ (figure \ref{fig:snapshots_local}C) . 
This cluster represents effectively an absorbing obstacle for the remaining mobile structures and leads eventually to a break-down of collective motion (figure \ref{fig:snapshots_local}D). However, at the end of the simulation run at $t\approx40000$ we observe again the onset of collective motion (figure \ref{fig:snapshots_local}E). Clearly, the bistable local dynamics predicted by the mean-field theory, leads to complex spatial and temporal dynamics for local coupling. Please note, that the time scales governing the large scale dynamics are in the order $\sim 10^4$. This is five orders of magnitude larger then the interaction time scale $\mu^{-1}=0.1$ which governs the microscopic dynamics. This large spread in the relevant time scales makes this system difficult to analyze by standard numerical methods and may require novel computational techniques, as for example the equation-free approach \cite{gear_equation-free_2003,kevrekidis_equationfree:_2004,kolpas_coarse-grained_2007}.

In the context of local coupling, we should mention recent studies on collective dynamics in different variants of the Vicsek model \cite{mishra_collective_2012,farrell_pattern_2012} and in a run-and-tumble model \cite{tailleur_statistical_2008}. In these models the velocity of individuals was assumed to be directly a function of the local density of individuals \cite{tailleur_statistical_2008,farrell_pattern_2012}, or of the local order parameter \cite{mishra_collective_2012}. These models show qualitatively similar phase-separation dynamics, resembling the behaviour observed here for  local coupling. However, in our model we do not explicitly include a dependency of the individual speed on the mean-field velocity and density, but it emerges naturally due to the interplay of the velocity-alignment force and the variable speed of individuals.

In conclusion, the ability of active Brownian particles to adjust their speed in response to interactions with the environment is certainly a more realistic, in comparison to the constant speed restriction in self-propelled particle models. However, this extension is far from being trivial. The mean-field approximation gives important insights in the emergent collective behaviour of the system and demonstrates clearly how the type of fluctuations can have a dramatic impact on the large scale collective dynamics.

%%%%%%%%%%%%%%%%%%%%%%%%%%%%%%%%%%%%%%%%%%%%%%%%%%%%%%%%%%%%%%%%%%%%%%%%%%%%%%
	
%%%%%%%%%%%%%%%%%%%%%%%%%%%%%%%%%%%%%%%%%%%%%%%%%%%%%%%%%%%%%%%%%%%%%%%%%%%%%%
%%								APPENDIX
%
\appendix

\section{Derivation of the moment equation}
\label{sec:app:derivation_of_moment_eqn}
The general evolution equation \eqnref{eqn:KT:moment:equation} for the moments $M^{(mn)}$ is derived in this section. Therefore it is exploited, that the dynamics of the active Brownian particle gas, which is determined by four stochastic differential equations \eqnref{eqn:KT:NU:eqn4}, can equally be described by the corresponding Fokker-Planck equation, which describes the dynamics of the total probability density function $P \left (\bi{r}_1,...,\bi{r}_N,\bi{v}_1,...,\bi{v}_N,t \right)$. The Stratonovich interpretation of the stochastic differential equation is used, which yields: 
	\begin{eqnarray}
		\label{Aeqn:KT:FFPeqn}
		 \fl \frac{\partial P}{\partial t} = \sum_{i=1}^N & \left \{-\nabla_{\bi{r}_i} \left ( \bi{v}_i \, P \right ) - \nabla_{\bi{v}_i} \left [ \left ( \frac{\bi{v}_i}{\abs{\bi{v}_i}} - \bi{v}_i + \mu \left ( \bi{u}_{A_i} - \bi{v}_i \right ) - D_{\varphi} \left ( \frac{\bi{v}_{i}}{\; \abs{\bi{v}_i}^2} \right )  \right ) P \right ]   \right. \nonumber\\
	 \fl &+ \, \frac{\partial^2}{\partial \left (\bi{e}_x \bi{v}_i\right)^2} \left [ \frac{D_v \left (\bi{e}_x \bi{v}_i\right)^2 + D_{\varphi}  \left (\bi{e}_y \bi{v}_i\right)^2}{\abs{\bi{v}_i}^2} \, P \right ] \nonumber \\
	 \fl &+\, \frac{\partial^2}{\partial \left (\bi{e}_y \bi{v}_i\right)^2} \left [ \frac{D_v  \left (\bi{e}_y \bi{v}_i\right)^2 + D_{\varphi} \left (\bi{e}_x \bi{v}_i\right)^2}{\abs{\bi{v}_i}^2} \, P \right ]  \nonumber \\
	 \fl &+ \left. \frac{\partial^2}{\partial \left (\bi{e}_x \bi{v}_i\right) \partial  \left (\bi{e}_y \bi{v}_i\right)} \left [ 2 \,  \frac{\left (\bi{e}_x \bi{v}_i\right)  \left (\bi{e}_y \bi{v}_i\right)}{\abs{\bi{v}_i}^2} \left ( D_v - D_{\varphi} \right ) P \right ]  + \Delta_{\bi{v}_i} \left ( D_{E} \, P \right ) \right \} \, \mbox{. }
	\end{eqnarray}
The average velocity $\bi{u}_{A_i}$ around the $i^{\mbox{\footnotesize{th}}}$ particle is given by the integral 
	\begin{eqnarray}
	\label{Aeqn:u1}
		\fl \bi{u}_{A_i} = \frac{1}{N} \, \sum_{j=1}^{N} \, \frac{ \int_{A_{i}} \rmd^2 \bi{r}_1...\rmd^2\bi{r}_N \int \rmd^2 \bi{v}_1...\rmd^2\bi{v}_N \, \bi{v}_j \, P\left (\bi{r}_1,...,\bi{r}_N,\bi{v}_1,...,\bi{v}_N,t \right) }{ \int_{A_{i}} \rmd^2 \bi{r}_1...\rmd^2\bi{r}_N \int \rmd^2 \bi{v}_1...\rmd^2\bi{v}_N \, P\left (\bi{r}_1,...,\bi{r}_N,\bi{v}_1,...,\bi{v}_N,t \right) } \, \mbox{, } 
	\end{eqnarray}
hence \eref{Aeqn:KT:FFPeqn} is a nonlinear Fokker-Planck equation. 
%The variables of integration denote $\rmd^2 \bi{r}_j=\rmd^2 x_j \rmd^2 y_j$ and $\rmd^2 \bi{v}_j = \rmd^2 v_{j,x} \rmd^2 v_{j,y}$ respectively. 
The interaction region of the $i^{\mbox{\footnotesize{th}}}$ particle is denoted by $A_{i}$. 

In the main text, the one-particle probability density function $p(\bi{r},\bi{v},t)$ is introduced. Its dynamics can easily be derived from \eref{Aeqn:KT:FFPeqn} by integrating over all particle positions and velocities except one, arbitrary chosen particle as well as using the mean field approximation
%Let $p\left (\bi{r}_j,\bi{v}_j,t \right )\rmd \bi{r}_j \rmd \bi{v}_j$ be the probability to find the $j^{\mbox{\footnotesize{th}}}$ particle in the velocity interval $\left [\bi{v}_j,\bi{v}_j+\rmd \bi{v}_j \right ]$ and within the spatial region $\left [\bi{r}_j,\bi{r}_j + \rmd \bi{r}_j \right ]$. The dynamics of the one-particle probability density function $p\left (\bi{r}_j,\bi{v}_j,t \right )$ can easily be derived from \eref{Aeqn:KT:FFPeqn} by integrating over all particle positions and velocities except the $j^{\mbox{\footnotesize{th}}}$ as well as using the mean field approximation
	\begin{eqnarray}
		\label{Aeqn:KT:approx:u}
			\bi{u}_{A}\left ( \bi{r},t\right ) & \approx \frac{\int_{A} \rmd^2 \bi{r} \int \rmd^2 \bi{v} \; \bi{v} \,  p(\bi{r},\bi{v},t) }{\int_{A} \rmd^2 \bi{r} \int \rmd^2 \bi{v} \; p(\bi{r},\bi{v},t) } \, \mbox{. }
	\end{eqnarray}
This assumption certainly holds for high particle densities and leads to an effective single-particle description:
%The dynamics of the single-particle probability density function $p\left (\bi{r},\bi{v},t\right )$ reads
	\begin{eqnarray}
		\label{Aeqn:KT:espFPE}
		 \fl \frac{\partial p}{\partial t} =& - \nabla_{\bi{r}} \left ( \bi{v} \, p \right ) - \nabla_{\bi{v}} \left [\left (\frac{\bi{v}}{\abs{\bi{v}}} - \bi{v}\right ) p + \mu \left (\bi{u}_{A} - \bi{v} \right ) p - D_{\varphi} \left (\frac{\bi{v}}{\; \abs{\bi{v}}^2} \right) p \right ] \nonumber  \\
		 \fl & + \, \frac{\partial^2}{\partial v_{x}^2} \left ( \frac{D_v v_{x}^2 + D_{\varphi} v_{y}^2}{\abs{\bi{v}}^2} \, p \right ) + \frac{\partial^2}{\partial v_{y}^2} \left ( \frac{D_v v_{y}^2 + D_{\varphi} v_{x}^2}{\abs{\bi{v}}^2} \,  p \right )  \nonumber \\
		 \fl & + \, \frac{\partial^2}{\partial v_{x} \partial v_{y}} \left [ 2 \,  \frac{v_{x} v_{y}}{\abs{\bi{v}}^2} \left ( D_v - D_{\varphi} \right ) p  \right ] + \Delta_{\bi{v}} \left ( D_{E} \, p \right ) \, \mbox{. }
	\end{eqnarray}	
Multiplying the moments of the velocity-components 
	\begin{equation}
		\label{Aeqn:KT:moment:definition}
		M^{(mn)} = \mean{v_x^m v_y^n} = \frac{1}{\rho(\bi{r},t)} \int \rmd^2 \bi{v} \; v_x^m v_y^n \; p(\bi{r},\bi{v},t)  
	\end{equation}
and the spatial density
	\begin{eqnarray}
		\label{Aeqn:KT:local density}
		M^{(00)}= \rho (\bi{r},t) = \int \rmd^2 \bi{v} \; p(\bi{r},\bi{v},t) 
	\end{eqnarray}
yields 
	\begin{eqnarray}
		\label{Aeqn:KT:ansatz:moment:expansion}
		\frac{\partial }{\partial t} \left ( \rho \, M^{(mn)} \right ) = \int \rmd^2 \bi{v} \; v_x^m v_y^n \; \frac{\partial p}{\partial t}  \, \mbox{. }
	\end{eqnarray}	
We may rewrite the integral on the right hand side of \eref{Aeqn:KT:ansatz:moment:expansion} by inserting the Fokker-Planck equation \eref{Aeqn:KT:espFPE}
	\begin{eqnarray}
		\fl \frac{\partial }{\partial t} \left ( \rho \, M^{(mn)} \right ) = &- \int \rmd^2 \bi{v} \; v_x^m v_y^n \; \frac{\partial}{\partial x} \left ( v_x \, p \right ) - \int \rmd^2 \bi{v} \; v_x^m v_y^n \; \frac{\partial}{\partial y} \left ( v_{y} \, p \right )  \nonumber \\
		\fl & - \int \rmd^2 \bi{v} \; v_x^m v_y^n \; \frac{\partial}{\partial v_x} \left [ \left (\abs{\bi{v}}^{-1} -1 \right )v_x \, p + \mu \left ( u_{A ,{x}} -v_x \right ) p - D_{\varphi} \left ( \frac{v_x}{\; \abs{\bi{v}}^2} \right) p \right ] \nonumber \\
		\fl & - \int \rmd^2 \bi{v} \; v_x^m v_y^n \; \frac{\partial}{\partial v_y} \left [ \left (\abs{\bi{v}}^{-1} -1 \right )v_y \, p + \mu \left ( u_{A ,{y}} - v_y \right ) p  - D_{\varphi} \left ( \frac{v_y}{\; \abs{\bi{v}}^2} \right) p \right ] \nonumber \\
		\fl & + \int \rmd^2 \bi{v} \; v_x^m v_y^n \; \frac{\partial^2}{\partial v_x^2} \left ( \frac{D_v v_{x}^2 + D_{\varphi} v_{y}^2}{\abs{\bi{v}}^2} \, p \right ) \nonumber \\
		\fl & + \int \rmd^2 \bi{v} \; v_x^m v_y^n \; \frac{\partial^2}{\partial v_{y}^2} \left ( \frac{D_v v_{y}^2 + D_{\varphi} v_{x}^2}{\abs{\bi{v}}^2} \, p \right ) \nonumber \\
		\fl & + \int \rmd^2 \bi{v} \; v_x^m v_y^n \; \frac{\partial^2}{\partial v_{x} \partial v_{y}} \left [ 2 \,  \frac{v_{x} v_{y}}{\abs{\bi{v}}^2} \left ( D_v - D_{\varphi} \right ) p \right ] \nonumber \\
		\fl & + \int \rmd^2 \bi{v} \; v_x^m v_y^n \; \frac{\partial^2}{\partial v_{x}^2} \left ( D_E \, p \right ) +  \int \rmd^2 \bi{v} \; v_x^m v_y^n \; \frac{\partial^2}{\partial v_{y}^2} \left ( D_E \, p \right )\ .  
	\end{eqnarray}
Integrals containing a derivative with respect to $v_{x}$ and $v_y$, respectively, are integrated by parts and rewritten using the moment definition \eref{Aeqn:KT:moment:definition}. Furthermore it is assumed that $p\left (\bi{r},\bi{v},t \right )$ decreases fast enough, so that
	\begin{eqnarray}
		\label{Aeqn:KT:vanishing_p}
		\lim_{v_{x} \rightarrow \infty } \left [ \lim_{v_y \to \infty} \left (  v_x^m v_y^n \, p(\bi{r},\bi{v},t ) \right ) \right ]= 0 
	\end{eqnarray}
holds. In this way, partial differential equations for the moments (macroscopic variables) of the probability density function are derived directly from the microscopic Langevin description \eref{eqn:ABM:active:particle:eqn}. The general evolution equation for the moment $M^{(mn)}$ reads 
	\begin{eqnarray}
		\label{Aeqn:KT:moment:equation}
		\fl \frac{\partial }{\partial t} \left ( \rho \, \mean{v_x^m v_y^n} \right ) = 
		    & - \frac{\partial}{\partial x} \left (\rho \mean{v_x^{m+1}v_y^n} \right ) - \frac{\partial }{\partial y} \left( \rho \mean{v_x^m v_{y}^{n+1}} \right) 
		\nonumber \\
		\fl & + \, (m+n) \rho \left[ \mean{\frac{v_x^m v_y^n}{\abs{\bi{v}}}} - \mean{v_x^m v_y^n} - D_\varphi \mean{\frac{v_x^m v_y^n}{\abs{\bi{v}}^2}} \right] 
		\nonumber \\
		\fl & + \mu \rho \left[ m\, u_{A,x}\mean{v_x^{m-1}v_y^n} + n\, u_{A,y}\mean{v_x^{m}v_y^{n-1}} - (m+n)\mean{v_x^m v_y^n} \right]  
		\nonumber \\
		\fl & + \,  m(m-1) \rho \left [D_{\varphi} \mean{\frac{v_x^{m-2} v_{y}^{n+2}}{\abs{\bi{v}}^2}} + D_v \mean{\frac{v_x^m v_y^n}{\,\abs{\bi{v}}^2}}  + D_E \mean{v_x^{m-2} v_y^{n}} \right ]
		\nonumber \\
		\fl & + \,  n(n-1) \rho \left [D_{\varphi} \mean{\frac{v_x^{m+2} v_{y}^{n-2}}{\abs{\bi{v}}^2}} + D_v \mean{\frac{v_x^m v_y^n}{\,\abs{\bi{v}}^2}}  + D_E \mean{v_x^{m}v_y^{n-2}} \right ]
		\nonumber \\
		\fl & +2 m n \rho \mean{\frac{v_x^{m} v_{y}^{n}}{\abs{\bi{v}}^2}}\left(D_v-D_\varphi \right) \, \mbox{. }
	\end{eqnarray}
It is shown in section \ref{sec:KT:ABM} how a closed set of hydrodynamic equations up to the second order can be derived from \eqnref{Aeqn:KT:moment:equation}. 

\section{Stochastic Heun method}
\label{sec:stoch:heun}

A system of stochastic differential equations with multiplicative Gaussian white noise ($\mean{\xi_i(t)}=0$, $\mean{\xi_i(t) \xi_j(t')}=\delta_{ij} \, \delta ( t-t')$),
	\begin{eqnarray}
		\dot{\bi{x}} = \bi{f} (\bi{x},t ) + \bi{g} ( \bi{x},t ) \, \boldsymbol{\xi}(t) \, \mbox{, }
	\end{eqnarray}
can numerically be integrated applying the stochastic Heun method. This integration scheme converges in quadratic mean to the solution of Langevin equations interpreted in the sense of Stratonovich \cite{ruemelin_numerical_1982} and reads as follows. 
	\Anumparts
		\begin{eqnarray}
			\fl \tilde{x}_i (t+\Delta t )  = x_i(t) && + f_i (\bi{x}(t),t) \Delta t + \sum_{j=1}^m \, g_{ij} ( \bi{x}(t),t) \sigma_j(t) \sqrt{\Delta t} \\
			\fl x_i(t+\Delta t) = x_i(t) &&+ \frac{1}{2} \left [ f_i (\bi{x}(t),t) + f_i (\tilde{\bi{x}}(t+\Delta t),t+\Delta t) \right] \Delta t \nonumber \\
			\fl & &+ \frac{1}{2} \sum_{j=1}^m \, \left [ g_{ij} ( \bi{x}(t),t) + g_{ij} ( \tilde{\bi{x}}(t+\Delta t),t+\Delta t) \right] \sigma_j(t) \sqrt{\Delta t}
		\end{eqnarray}
	\endAnumparts
For the integration, Gaussian random variables $\sigma_j$ with zero mean and variance one are used.
	\Anumparts
		\begin{eqnarray}
			\mean{\sigma_i(t)} 				&= \; 0 \\
			\mean{\sigma_i(t) \sigma_j(t)}	&= \; \delta_{ij}
		\end{eqnarray}
	\endAnumparts

\section*{References} 

\bibliographystyle{unsrt}	
\bibliography{Grossmann_et_al}

\begin{thebibliography}{10}

\bibitem{couzin_collective_2002}
I.~D. Couzin, J.~Krause, R.~James, G.~D. Ruxton, and N.~R. Franks.
\newblock Collective memory and spatial sorting in animal groups.
\newblock {\em Journal of Theoretical Biology}, 218(1):1--11, September 2002.

\bibitem{krause_living_2002}
J.~Krause and G.~D. Ruxton.
\newblock {\em Living in groups}.
\newblock Oxford University Press, 2002.

\bibitem{parrish_complexity_1999}
J.~K. Parrish and L.~{Edelstein-Keshet}.
\newblock Complexity, pattern, and evolutionary {Trade-Offs} in animal
  aggregation.
\newblock {\em Science}, 284(5411):99--101, April 1999.

\bibitem{vicsek_novel_1995}
T.~Vicsek, A.~Czir{\'o}k, E.~{Ben-Jacob}, I.~Cohen, and O.~Shochet.
\newblock Novel type of phase transisition in a system of {Self-Driven}
  particles.
\newblock {\em Physical Review Letters}, 75(6):1226--1229, 1995.

\bibitem{toner_long-range_1995}
J.~Toner and Y.~Tu.
\newblock {Long-Range} order in a {Two-Dimensional} dynamical {XY} model: How
  birds fly together.
\newblock {\em Physical Review Letters}, 75(23):4326, 1995.

\bibitem{toner_flocks_1998}
J.~Toner and Y.~Tu.
\newblock Flocks, herds, and schools: A quantitative theory of flocking.
\newblock {\em Physical Review E}, 58(4):4828--4858, 1998.

\bibitem{gregoire_onset_2004}
G.~Gr{\'e}goire and H.~Chat\'e.
\newblock Onset of collective and cohesive motion.
\newblock {\em Physical Review Letters}, 92(2):025702, 2004.

\bibitem{aldana_phase_2007}
M.~Aldana, V.~Dossetti, C.~Huepe, V.~M. Kenkre, and H.~Larralde.
\newblock Phase transitions in systems of {Self-Propelled} agents and related
  network models.
\newblock {\em Physical Review Letters}, 98(9):095702--4, 2007.

\bibitem{chate_collective_2008}
H.~Chat\'e, F.~Ginelli, G.~Gr{\'e}goire, and F.~Raynaud.
\newblock Collective motion of self-propelled particles interacting without
  cohesion.
\newblock {\em Physical Review E}, 77(4):046113, April 2008.

\bibitem{romanczuk_collective_2009}
P.~Romanczuk, I.~D. Couzin, and L.~{Schimansky-Geier}.
\newblock Collective motion due to individual escape and pursuit response.
\newblock {\em Physical Review Letters}, 102(1):010602--4, January 2009.

\bibitem{kudrolli_swarming_2008}
A.~Kudrolli, G.~Lumay, D.~Volfson, and L.~S. Tsimring.
\newblock Swarming and swirling in {Self-Propelled} polar granular rods.
\newblock {\em Physical Review Letters}, 100(5):058001, February 2008.

\bibitem{deseigne_collective_2010}
J.~Deseigne, O.~Dauchot, and H.~Chat\'e.
\newblock Collective motion of vibrated polar disks.
\newblock {\em Physical Review Letters}, 105(9):098001, 2010.

\bibitem{schaller_polar_2010}
V.~Schaller, C.~Weber, C.~Semmrich, E.~Frey, and A.~R. Bausch.
\newblock Polar patterns of driven filaments.
\newblock {\em Nature}, 467(7311):73--77, 2010.

\bibitem{peruani_collective_2012}
F.~Peruani, J.~Starru{\ss}, V.~Jakovljevic, L.~{S{\o}gaard-Andersen},
  A.~Deutsch, and M.~B\"ar.
\newblock Collective motion and nonequilibrium cluster formation in colonies of
  gliding bacteria.
\newblock {\em Physical Review Letters}, 108(9):098102, February 2012.

\bibitem{nagy_new_2007}
M.~Nagy, I.~Daruka, and T.~Vicsek.
\newblock New aspects of the continuous phase transition in the scalar noise
  model {(SNM)} of collective motion.
\newblock {\em Physica A: Statistical and Theoretical Physics}, 373:445--454,
  January 2007.

\bibitem{baglietto_nature_2009}
G.~Baglietto and E.~V. Albano.
\newblock Nature of the order-disorder transition in the vicsek model for the
  collective motion of self-propelled particles.
\newblock {\em Physical Review E}, 80(5):050103, November 2009.

\bibitem{bazazi_nutritional_2011}
S.~Bazazi, P.~Romanczuk, S.~Thomas, L.~{Schimansky-Geier}, J.~J. Hale, G.~A.
  Miller, G.~A. Sword, S.~J. Simpson, and I.~D. Couzin.
\newblock Nutritional state and collective motion: from individuals to mass
  migration.
\newblock {\em Proceedings of the Royal Society B: Biological Sciences},
  278(1704):356 --363, February 2011.

\bibitem{peruani_self-propelled_2007}
F.~Peruani and L.~G. Morelli.
\newblock {Self-Propelled} particles with fluctuating speed and direction of
  motion in two dimensions.
\newblock {\em Physical Review Letters}, 99(1):010602, July 2007.

\bibitem{stroembom_collective_2011}
D.~Str\"ombom.
\newblock Collective motion from local attraction.
\newblock {\em Journal of Theoretical Biology}, 283(1):145--151, August 2011.

\bibitem{lukeman_inferring_2010}
R.~Lukeman, {Y.-X.} Li, and L.~{Edelstein-Keshet}.
\newblock Inferring individual rules from collective behavior.
\newblock {\em Proceedings of the National Academy of Sciences},
  107(28):12576--12580, 2010.

\bibitem{katz_inferring_2011}
Y.~Katz, K.~Tunstr{\o}m, C.~C. Ioannou, C.~Huepe, and I.~D. Couzin.
\newblock Inferring the structure and dynamics of interactions in schooling
  fish.
\newblock {\em Proceedings of the National Academy of Sciences}, July 2011.

\bibitem{herbert-read_inferring_2011}
J.~E. {Herbert-Read}, A.~Perna, R.~P. Mann, T.~M. Schaerf, D.~J.~T. Sumpter,
  and A.~J.~W. Ward.
\newblock Inferring the rules of interaction of shoaling fish.
\newblock {\em Proceedings of the National Academy of Sciences}, November 2011.

\bibitem{simha_hydrodynamic_2002}
R.~A. Simha and S.~Ramaswamy.
\newblock Hydrodynamic fluctuations and instabilities in ordered suspensions of
  {Self-Propelled} particles.
\newblock {\em Physical Review Letters}, 89(5):058101, July 2002.

\bibitem{schweitzer_complex_1998}
F.~Schweitzer, W.~Ebeling, and B.~Tilch.
\newblock Complex motion of brownian particles with energy depots.
\newblock {\em Physical Review Letters}, 80(23):5044, June 1998.

\bibitem{erdmann_brownian_2000}
U.~Erdmann, W.~Ebeling, L.~{Schimansky-Geier}, and F.~Schweitzer.
\newblock Brownian particles far from equilibrium.
\newblock {\em The European Physical Journal B - Condensed Matter and Complex
  Systems}, 15(1):105--113, April 2000.

\bibitem{schweitzer_brownian_2003}
F.~Schweitzer.
\newblock {\em Brownian Agents and Active Particles: Collective Dynamics in the
  Natural and Social Sciences}.
\newblock Springer, 1 edition, August 2003.

\bibitem{garcia_testing_2011}
V.~Garcia, M.~Birbaumer, and F.~Schweitzer.
\newblock Testing an agent-based model of bacterial cell motility: How nutrient
  concentration affects speed distribution.
\newblock {\em The European Physical Journal {B-Condensed} Matter and Complex
  Systems}, 82(3):235–244, 2011.

\bibitem{romanczuk_brownian_2011}
P.~Romanczuk and L.~{Schimansky-Geier}.
\newblock Brownian motion with active fluctuations.
\newblock {\em Physical Review Letters}, 106(23):230601, June 2011.

\bibitem{chate_comment_2007}
H.~Chat\'e, F.~Ginelli, and G.~Gr{\'e}goire.
\newblock Comment on {“Phase} transitions in systems of {Self-Propelled}
  agents and related network models”.
\newblock {\em Physical Review Letters}, 99(22):229601, November 2007.

\bibitem{schienbein_langevin_1993}
M.~Schienbein and H.~Gruler.
\newblock Langevin equation, {Fokker-Planck} equation and cell migration.
\newblock {\em Bulletin of Mathematical Biology}, 55(3):585--608, May 1993.

\bibitem{condat_randomly_2005}
C.~A. Condat, J.~J{\"a}ckle, and S.~A. Mench{\'o}n.
\newblock Randomly curved runs interrupted by tumbling: A model for bacterial
  motion.
\newblock {\em Physical Review E}, 72(2):021909, 2005.

\bibitem{kuramoto_self-entrainment_1975}
Y.~Kuramoto.
\newblock Self-entrainment of a population of coupled non-linear oscillators.
\newblock In H.~Araki, editor, {\em International Symposium on Mathematical
  Problems in Theoretical Physics}, volume~39, pages 420--422.
  {Springer-Verlag}, {Berlin/Heidelberg}, 1975.

\bibitem{paley_oscillator_2007}
D.~A. Paley, N.~E. Leonard, R.~Sepulchre, and D.~Grunbaum.
\newblock Oscillator models and collective motion.
\newblock {\em {IEEE} {CONTROL} {SYSTEMS} {MAGAZINE}}, 2007.

\bibitem{birnir_ode_2007}
Bj{\"o}rn Birnir.
\newblock An {ODE} model of the motion of pelagic fish.
\newblock {\em Journal of Statistical Physics}, 128(1-2):535--568, February
  2007.

\bibitem{peruani_mean-field_2008}
F.~Peruani, A.~Deutsch, and M.~B\"ar.
\newblock A mean-field theory for self-propelled particles interacting by
  velocity alignment mechanisms.
\newblock {\em The European Physical Journal Special Topics}, 157(1):111--122,
  April 2008.

\bibitem{czirok_formation_1996}
Andr{\'a}s Czir{\'o}k, Eshel {Ben-Jacob}, Inon Cohen, and Tam{\'a}s Vicsek.
\newblock Formation of complex bacterial colonies via self-generated vortices.
\newblock {\em Physical Review E}, 54(2):1791--1801, 1996.

\bibitem{niwa_self-organizing_1994}
{H.-S.} Niwa.
\newblock Self-organizing dynamic model of fish schooling.
\newblock {\em Journal of Theoretical Biology}, 171(2):123--136, November 1994.

\bibitem{bertin_boltzmann_2006}
E.~Bertin, M.~Droz, and G.~Gr{\'e}goire.
\newblock Boltzmann and hydrodynamic description for self-propelled particles.
\newblock {\em Physical Review E {(Statistical}, Nonlinear, and Soft Matter
  Physics)}, 74(2):022101--4, 2006.

\bibitem{bertin_microscopic_2009}
E.~Bertin, M.~Droz, and G.~Gr{\'e}goire.
\newblock Microscopic derivation of hydrodynamic equations for self-propelled
  particles.
\newblock {\em Journal of Physics A: Mathematical and Theoretical}, 42:445001,
  2009.

\bibitem{ihle_kinetic_2011}
T.~Ihle.
\newblock Kinetic theory of flocking: Derivation of hydrodynamic equations.
\newblock {\em Physical Review E}, 83(3):030901, March 2011.

\bibitem{baskaran_hydrodynamics_2008}
A.~Baskaran and M.~C. Marchetti.
\newblock Hydrodynamics of self-propelled hard rods.
\newblock {\em Physical Review E}, 77(1):011920, January 2008.

\bibitem{mishra_fluctuations_2010}
S.~Mishra, A.~Baskaran, and M.~C. Marchetti.
\newblock Fluctuations and pattern formation in self-propelled particles.
\newblock {\em Physical Review E}, 81(6):061916, June 2010.

\bibitem{riethmueller_langevin_1997}
T.~Riethm\"uller, L.~{Schimansky-Geier}, D.~Rosenkranz, and T.~P\"oschel.
\newblock Langevin equation approach to granular flow in a narrow pipe.
\newblock {\em Journal of Statistical Physics}, 86(1):421--430, January 1997.

\bibitem{romanczuk_collective_2010}
P.~Romanczuk and U.~Erdmann.
\newblock Collective motion of active brownian particles in one dimension.
\newblock {\em The European Physical Journal Special Topics}, 187(1):127--134,
  October 2010.

\bibitem{romanczuk_mean-field_2011}
P.~Romanczuk and L.~{Schimansky-Geier}.
\newblock Mean-field theory of collective motion due to velocity alignment.
\newblock {\em Ecological Complexity}, 2011.

\bibitem{csahok_hydrodynamics_1997}
Zolt{\'a}n Csah{\'o}k and Andr{\'a}s Czir{\'o}k.
\newblock Hydrodynamics of bacterial motion.
\newblock {\em Physica A}, 243(3-4):304--318, 1997.

\bibitem{ruemelin_numerical_1982}
W.~R\"umelin.
\newblock Numerical treatment of stochastic differential equations.
\newblock {\em {SIAM} Journal on Numerical Analysis}, 19(3):604--613, June
  1982.

\bibitem{strefler_swarming_2008}
J.~Strefler, U.~Erdmann, and L.~{Schimansky-Geier}.
\newblock Swarming in three dimensions.
\newblock {\em Physical Review E}, 78(3):031927, 2008.

\bibitem{mirollo_amplitude_1990}
R.~E. Mirollo and S.~H. Strogatz.
\newblock Amplitude death in an array of limit-cycle oscillators.
\newblock {\em Journal of Statistical Physics}, 60(1-2):245--262, July 1990.

\bibitem{lee_fluctuation-induced_2010}
C.~F. Lee.
\newblock Fluctuation-induced collective motion: A single-particle density
  analysis.
\newblock {\em Physical Review E}, 81(3):031125, 2010.

\bibitem{gear_equation-free_2003}
C.~William Gear, J.~M. Hyman, P.~G. Kevrekidis, I.~G. Kevrekidis, Olof Runborg,
  and C.~Theodoropoulos.
\newblock {Equation-Free}, {Coarse-Grained} multiscale computation: Enabling
  microscopic simulators to perform {System-Level} analysis.
\newblock {\em Communications in Mathematical Sciences}, 1(4):715--762,
  December 2003.
\newblock Mathematical Reviews number {(MathSciNet):} {MR2041455}.

\bibitem{kevrekidis_equationfree:_2004}
I.~G. Kevrekidis, C.~W. Gear, and G.~Hummer.
\newblock Equation‐free: The computer‐aided analysis of complex multiscale
  systems.
\newblock {\em {AIChE} Journal}, 50(7):1346--1355, July 2004.

\bibitem{kolpas_coarse-grained_2007}
A.~Kolpas, J.~Moehlis, and I.~G. Kevrekidis.
\newblock Coarse-grained analysis of stochasticity-induced switching between
  collective motion states.
\newblock {\em Proceedings of the National Academy of Sciences}, 104(14):5931
  --5935, April 2007.

\bibitem{mishra_collective_2012}
S.~Mishra, K.~Tunstr{\o}m, I.~D. Couzin, and C.~Huepe.
\newblock Collective dynamics of self-propelled particles with variable speed.
\newblock {\em {arXiv:1202.3495}}, February 2012.

\bibitem{farrell_pattern_2012}
F.~D.~C Farrell, J.~Tailleur, D.~Marenduzzo, and M.~C Marchetti.
\newblock Pattern formation in self-propelled particles with density-dependent
  motility.
\newblock {\em {arXiv:1202.0749}}, February 2012.

\bibitem{tailleur_statistical_2008}
J.~Tailleur and M.~E. Cates.
\newblock Statistical mechanics of interacting {Run-and-Tumble} bacteria.
\newblock {\em Physical Review Letters}, 100(21):218103, May 2008.

\end{thebibliography}

%\bibliography{Grossmann_et_al_paper_style}		% expects file "myrefs.bib"
%\bibliographystyle{Grossmann_et_al_bibstyle}	% (uses file "plain.bst")

\end{document}